\documentclass[pra]{revtex4-1}

\usepackage{graphicx}
\usepackage{amssymb}
\usepackage{epstopdf}
\usepackage{amsfonts}
\usepackage{amsmath}
\usepackage{bm}
\usepackage{mathrsfs}
\usepackage{subfigure}

\begin{document}

\title{Nonequilibrium Atom-Dielectric Forces Mediated by a Quantum Field}

\author{Ryan O. Behunin$^{1, 2, 3}$ and Bei-Lok Hu$^{1,2}$}

\affiliation{$^1$Maryland Center for Fundamental Physics, Department
of Physics,
\\ $^2$Joint Quantum Institute, University of Maryland,
College Park, Maryland, 20742, USA\\
$^3$ Theoretical Division, Los Alamos National Laboratory, Los Alamos, NM 87545, USA}

\date{November 1, 2010}

\begin{abstract}
In this paper we give a first principles microphysics derivation of the nonequilibrium forces between an atom, treated as a three dimensional harmonic oscillator, and a  bulk dielectric medium modeled as a continuous lattice of oscillators coupled to a reservoir. We assume no direct interaction between the atom and the medium but there exist mutual influences transmitted via a common electromagnetic field. By employing concepts and techniques of open quantum systems we introduce coarse-graining to the physical variables -- the medium, the quantum field and the atom's internal degrees of freedom, in that order --  to extract their averaged effects from the lowest tier progressively to the top tier.   The first tier of coarse-graining provides the averaged effect of the medium upon the field, quantified by a complex permittivity (in the frequency domain) describing the response of the dielectric to the field in addition to its back action on the field through a stochastic forcing term.  The last tier of coarse-graining over the atom's internal degrees of freedom   results in an equation of motion for the atom's center of mass from which we can derive the force on the atom. Our nonequilibrium formulation provides a fully dynamical description of the atom's motion including back action effects from all other relevant variables concerned.  In the long-time limit we recover the known results for the atom-dielectric force when the combined system is in equilibrium or in a nonequilibrium stationary state.
\end{abstract}

\maketitle

 %%%%%%%%%%%%%%%%%%%%%%%%%%%%%%%
\section{Introduction}
 %%%%%%%%%%%%%%%%%%%%%%%%%%%%%%%

%It is a remarkable fact that the dominant force between charge neutral atoms arises from quantum fluctuations.
%Particularly, fluctuations of the charge density about the nucleus,  arising from two physically distinct processes, give rise to instantaneous multipole moments
%First, microscopic degrees of freedom treated quantum mechanically, like the dipole moments of atoms, possess \textit{intrinsic} fluctuations,
%and second, intrinsic fluctuations of the quantized electromagnetic field \textit{induce} instantaneous dipole moments in atoms.
%In this way at zero temperature (and large separations) one can view the force between neutral atoms as a manifestation of the quantum vacuum.

%Research is this area is justified by the experimental needs in; the manipulation of trapped atoms [cite], the design and operation of micromechanical devices \cite{Chan}, measuring the effects of non-Newtonian forces \cite{nonnewt}, and permitting the investigation of quantum field theory in the laboratory without the need for accelerators.

It is a remarkable fact that the dominant force between neutral atoms at short distances (e.g.  the London force between two hydrogen atoms dominates at distances $\lesssim$ 1mm)
arises from quantum fluctuations.  Indeed, at zero temperature for
distances  larger than the wavelength associated with an atom's first
optical resonance the interaction originates from vacuum
fluctuations. The last decade saw intensified research in this area
of fluctuation-generated forces between atoms and between an atom and
a conducting or dielectric surface. This was brought about by
increased high-precision capability in the manipulation of
trapped atoms in cavities and optical lattices \cite{cavity1,cavity2},
superconductivity experiments \cite{supercond} and the design and
operation of nano- and micro-electromechanical devices \cite{nem,Chan01a,Chan01b}, amongst
others. These advances which hold the promise of ushering in a new era
of quantum engineering also made possible a wide range of theoretical
inquires, such as the measurements of non-Newtonian forces
\cite{nonnewt1,nonnewt2,nonnewt3}, utilization of experimental systems out of
thermal equilibrium \cite{Cornell,ReyHu}, and the investigation of quantum
field theoretical effects in tabletop experiments. The expanded capability of ultrafast, high intensity
lasers and high-precision manipulation techniques on atoms make it
possible to observe real time processes which in turn demands a new
theoretical framework to treat systems far from equilibrium. In
our research program which began with the two earlier papers
\cite{RH09,RH10} and continues with the present work we apply the
methods of nonequilibrium quantum field theory \cite{NEqQFT} within
the open quantum system \cite{qos} conceptual framework to tackle these
problems. This method can provide a fully dynamical description of
(non-stationary) systems far from equilibrium under the influence of
various environments, or acted upon by different noises, going beyond
the traditional mean field or linear response treatments.

The state-of-the-art as far as we can discern from the literature is
the so-called  macroscopic quantum electrodynamics (MQED)
which describes electromagnetic field fluctuations in a lossy medium.  In order for the system to remain in
thermal equilibrium the energy absorbed by the dielectric medium is
compensated for by a stochastic forcing term which is added by hand
in a manner consistent with the fluctuation-dissipation relation
(FDR) \cite{FDR}. Previous works employing MQED (or its variants)
 \cite{McLachlan63a, McLachlan63, Wylie84, Wylie85, APS05, Scheel&Buhmann,
Ant_JPhys, APSS08} (to name a few) have been skillfully employed to
study atom-surface forces (or spontaneous emission) for a plethora of experimental setups. The
key limitation of this technique is that it requires the system to be in a steady-state, or
at least in local thermal equilibrium in order to apply the FDR. The
method we use reproduces these earlier results for systems under equilibrium or
nonequilibrium steady-state conditions, but is capable of treating
fully nonequilibrium conditions including the effects of relaxation,
 and arbitrary atomic motion
including full back-action from the field and the medium in a
self-consistent manner. The FDR emerges from our model under steady-state 
conditions as a consequence of our energy conserving microscopic formulation 
in distinction to being \textit{required} as in the formulation of Rytov's theory \cite{Rytov59}.

The techniques of MQED have been adapted to describe nonequilibrium
atom-surface forces for two scenarios.
 In \cite{Buhmann09} a full treatment of the interactions between the  
 body-assisted field \textit{in equilibrium} and an atom in an arbitrary 
 state was undertaken. There, it is the time-devolpment of the atomic state that  
 characterizes the nonequilibrium behavior.
In distinction Antezza et. al. \cite{APS05} have successfully
applied a nonequilibrium generalization of MQED  to
describe atom-surface forces when the field and surface are not in
global thermodynamic equilibrium yet remain stationary (see \cite{Ellingsen10} for 
a unified treatment of both scenarios).  
 This generalization relies
crucially on the validity of a `local source hypothesis', which
assumes no spatial correlation of the fluctuating polarization that
drives the field in MQED. This \textit{ad hoc} hypothesis is equivalent to
ignoring interactions among the micro constituents of the dielectric
medium. When the temperature of the body is much higher than the
interaction energy among the medium's micro elements we believe that
the `local source hypothesis' should be an excellent approximation,
and that the permittivity can safely be assumed to be local in space. But at
low temperatures or when the coherence length of fluctuations in the
medium become large this approximation is expected to break down and
new techniques are needed to probe the fully nonequilbrium regimes.

%Our methods based on nonequilibrium quantum field theory provide such a fully dynamical description and are valid for general interacting media.

A key challenge in this endeavor is to understand the effects of
dissipation \cite{Rosa2010} on the field. Introducing a complex
permittivity to the theory by hand provides dissipation but the energy of the field is not conserved, and so
the field cannot be quantized in terms of energy eigenmodes. 
A first step was taken in \cite{CarMan71, Eberlein06} who considered
quantization of the field in a dielectric half-space for the case of
a real and frequency-independent permittivity. For real permittivity
there is no dissipation and the field can readily be quantized but the permittivity
violates the Kramers-Kronig relations \cite{Kramers,Kronig} and so
engenders acausal response.

Our ultimate aim in this paper is a first-principles derivation of the
atom-surface force based on the microphysics 
%We will account for the dissipation of the field by beginning with an action that describes the microscopic physics
of the dielectric medium, the quantum field, and the atom. At the micro-level the \textit{total combined system} is energy conserving. Dissipative effects arise at the macroscopic level after coarse-graining the detailed information in one or more subsystems.
We employ the model of Huttner and Barnett \cite{HB92}  who described the micro-elements of a dielectric material by a continuous lattice of harmonic oscillators  (from now on we'll refer to it as the matter field) coupled to a reservoir  (see Appendix A of \cite{Rahi} for the matter described in terms of a fermionic theory). The use of a reservoir provides a microscopic method for introducing dissipation into the remaining degrees of freedom in the total system.
%(though in our formulation a reservoir introduced specifically for this purpose is not necessary)  
For such a model composed of field + dielectric + reservoir the problem can be solved
exactly by Fano's diagonalization \cite{Fano} when the coupling between respective components is bilinear.
In distinction to that work, we are not particularly interested in
the microscopic details of the dielectric but only need to capture
the averaged effect of the medium on the quantum field and the atom. 
By invoking the concepts and techniques for open quantum systems we 
\textit{coarse-grain} the medium by tracing over the dielectric variables 
leading to a complex permittivity (in the frequency domain) which accounts
for the dielectric's response to the field, and the absorption and
emission of energy. In addition, the fluctuations of the oscillators which make 
up the dielectric manifest as a stochastic polarization that drives the field, 
and when the system is in
equilibrium serve to balance the dissipative losses.
%, and only when these fluctuations are accounted for will the total energy of the composite system be conserved.
From this microscopic viewpoint we see that a complex permittivity,
quantifying material response to an external field, \textit{cannot}
be added to the theory freely unless the fluctuations of the same
degrees of freedom are accounted for. It can be shown that at this
level the semi-classical equations of motion for the field under the
influence of the medium at late times (\ref{SCEM}) take the exact same form as in
MQED (Lifshitz) theory when the medium is assumed to be in a thermal state.
The microscopic approach we take offers a unique
vantage point to see that the stochastic polarization driving the
field put in by hand in MQED without field quantization
actually arises from what is equivalent to the media's fluctuations.
One could interpret the field fluctuations in MQED as being
\textit{induced} by a fictitious matter field.

The approach we adopt is similar to  that of \cite{Bechler} where a
path integral formulation was used to derive the effective action
describing the medium- influenced dynamics of the electromagnetic
field. For the specific case of a dielectric half-space our results
can be compared with \cite{YG} who generalized the results of
Carniglia and Mandel \cite{CarMan71} to frequency- dependent and lossy permittivities.
%This was done by beginning from a microscopic formulation where was 
They calculated exactly the dielectric + field dynamics using the
Wiener-Hopf method and a sum over diagrams. We go beyond these
results by providing a fully nonequilibrium treatment and apply these results
to the atom-surface force.

The paper is organized as follows. In Sec. 2 we describe the
microscopic model. In Sec. 3 we find the medium altered
equations of motion for the quantum field. From there the permittivity of the medium is identified as
well as a stochastic force which accounts for the response and
fluctuations of the dielectric. In Sec. 4 and Sec. 5 successive layers of
coarse-graining are performed to obtain the dynamics of the atom's
trajectory. In Sec. 6. we consider the specific case of the atom-surface 
force in a dielectric half-space and compare with the known results 
from the literature. We adopt the Einstein summation convention and
 natural units throughout $\hbar = c = k_B =1$.

 %%%%%%%%%%%%%%%%%%%%%%%%%%%%%%%
\section{Microscopic Model}
 %%%%%%%%%%%%%%%%%%%%%%%%%%%%%%%

The action describing the entire system $S[ \vec{z}, \vec{Q}, A^\mu,
\vec{P}, \vec{X}_\nu ]$ is the sum of eight terms:

\begin{equation}
S[ \vec{z}, \vec{Q}, A^\mu, \vec{P}, \vec{X}_\nu] \equiv S_Z + S_Q + S_E + S_M + S_X+  S^{AF}_{int} + S^{PF}_{int} + S^{PX}_{int},
\end{equation}  with five free actions pertaining to the five dynamical variables and three interaction actions. 
Here 1) $S_Z$ is the action for the motion of the atom's center of mass with coordinate  $\vec{z}$ and total mass $M$ under the
influence of an external potential $V[\vec{z}(\lambda)]$:

\begin{equation}
\label{ }
S_Z[\vec{z}]=  \int_{t_i}^{t_f} dt  \ \bigg[  \frac{M}{2} {\dot{\vec{z}}}^2(t)-V[\vec{z}(t)] \bigg]
\end{equation}
where $t$ is the atom's worldline parameter. 2) The internal
degrees of freedom of the atom are modeled by a three dimensional
harmonic oscillator with coordinate $\vec{Q}$ and natural frequency
$\Omega$ with action

\begin{equation}
\label{ }
S_Q[\vec{Q}]= \frac{\mu}{2}  \int_{t_i}^{t_f} dt  \ \left[ \dot{\vec{Q}}^2(t)-\Omega^2 \vec{Q}^2(t) \right] 
\end{equation} where $\mu$ is the oscillator's reduced mass. 3) The dynamics of
the free photon field is described by $S_E$ where `E' stands for the
electromagnetic field, 

\begin{equation}
\label{ }
S_E[A^\mu]= \frac{1}{4}\int d^4 x \  F^{\mu \nu} F_{\mu \nu}.
\end{equation} where $F_{\mu \nu}= \partial_\mu A_\nu-\partial_\nu A_\mu $ is the field strength tensor, with $A_\mu$ the photon field's vector potential and $\int d^4 x = \int_{t_i}^{t_f} dt \int d^3 x$.
4) The matter may be described by  a continuous lattice of harmonic oscillators with natural frequency $\omega$ which is meant to model the polarization of the medium, the coordinate of each oscillator is described by
the vector field, $\vec{P}$,  with action

\begin{equation}
\label{ }
S_M[\vec{P}]= \frac{1}{2}\int_V d^4 x[ \dot{\vec{P}}^2(x)-\omega^2 \vec{P}^2(x)].
\end{equation}
The subscript $V$ on the integration denotes that 
the spatial integration is restricted to the volume containing the matter.

5) Each oscillator comprising the matter field is coupled to a reservoir. The 
reservoir is composed of a collection of oscillators at each point 
with frequency dependent mass $I(\nu)$ and coordinates $\vec{X}_\nu$,
 with natural frequency
$\nu$
%One can view each point within the medium as being occupied by
%the coordinate of each represents its dipole moment which couples to the field.

\begin{equation}
\label{ }
S_X[\vec{X}_\nu]= \frac{1}{2}\int_V d^4 x \int d \nu \ I(\nu)[ \dot{\vec{X}}_\nu^2(x)-\nu^2 \vec{X}^2(x)].
\end{equation}

The interaction of these parties is specified by the three interaction actions. 
6) The interaction between
the internal degree of freedom (dof) of the atom and the field is
%will be of the general form of the dot product of  In the case of the atom we have

\begin{equation}
\label{ }
S^{AF}_{int}[A^\mu,  \vec{Q}, \vec{z}] = q \int d\lambda \  Q^i(\lambda) E_i(\lambda, \vec{z}(\lambda))
\end{equation}
where $q$ represents the electronic charge. For the case of the
matter the dipole moment of each oscillator is coupled with the local
electric field with a coupling $\Omega_P$

\begin{equation}
\label{ }
S^{PF}_{int}[A^\mu, \vec{P}] = \Omega_P  \int_V d^4 x \ P^i_\nu(x) E_i(x).
\end{equation}
For our model, which considers only the coupling of the 
electric field to the local polarization of the matter, there is no magnetic response
for the medium and so the permeability can assume its vacuum value, $\mu_o$, throughout.
 It should be noted that we have \textit{not}
included interactions among the elements of the dielectric which will provide 
a spatially local form for the permittivity in the macroscopic Maxwell's equations.

Finally, each oscillator composing the matter is coupled to a reservoir with frequency 
dependent `charge' $g(\nu)$ which will provide dissipation and noise

\begin{equation}
\label{ }
S^{PX}_{int}[ \vec{P}, \vec{X}_\nu] = \int_V d^4 x \int_0^\infty d\nu \ g(\nu) P_i (x) X^i_\nu(x).
\end{equation}

 %%%%%%%%%%%%%%%%%%%%%%%%%%%%%%%
\section{Field Equations in the presence of a Medium}
 %%%%%%%%%%%%%%%%%%%%%%%%%%%%%%%

%%%%% RESERVOIR REDUCTION %%%%%

In this section we will show how coarse-graining over the medium
degrees of freedom leads to a permittivity and a classical stochastic
source responsible for an additional induced component of field fluctuations.
Here, when we refer to medium we mean the combined reservoir + matter system. 
For now let us forget about the atom and field and focus entirely upon
the matter and reservoir. 

\subsection{Reservoir-Reduced Density Matrix}

Consider the time evolution of the density matrix describing the
medium (matter + reservoir system), $\hat{\rho}(t)=\hat{U}(t,t_i)
\hat{\rho}(t_i) \hat{U}^\dag(t,t_i)$ from some initial state at time $t_i$ 
to a final time $t$. By considering the matrix elements
in an appropriate basis we can express $\hat{\rho}(t)$ as a product
of path integrals

\begin{eqnarray}
\label{ }
\rho(\vec{P}_{ f}, \vec{P}_{ f}',\vec{X}_{\nu f}, \vec{X}_{\nu f}'; t)
= 
 \int^{\vec{P}_f, \vec{P}_f'}_{CTP} \mathcal{D} \vec{P}
 \
\exp \{ i(S_M[\vec{P}]-S_M[\vec{P}'] +S^{PF}_{int}[A^\mu, \vec{P}]-S^{PF}_{int}[A^{\mu'}, \vec{P}']) \}  
\nonumber \\
\times
\prod_\nu  \int^{\vec{X}_{ \nu f}, \vec{X}_{\nu f}'}_{CTP} \mathcal{D} \vec{X}_\nu \
\exp \{ i(S_X[\vec{X}_\nu]-S_X[\vec{X}_\nu'] +S^{PX}_{int}[ \vec{P}, \vec{X}_\nu ]-S^{PX}_{int}[\vec{P}', \vec{X}_\nu']) \} 
\end{eqnarray}
where the density matrix depends upon the electromagnetic field through the interaction $S^{PF}_{int}$ and $CTP$ stands for \textit{closed-time path}.
We will assume that the initial state of the total system factorizes allowing the convenient notation

\begin{eqnarray}
\int^{\vec{Y}_f, \vec{Y}_f'}_{CTP} \mathcal{D} \vec{Y} =  \int \mathcal{D} \vec{Y}_{ i}
\int \mathcal{D} \vec{Y}_{ i}'
\int_{\vec{Y}_{ i}}^{\vec{Y}_{ f}} \mathcal{D} \vec{Y}
\int_{\vec{Y}_{ i}'}^{\vec{Y}_{ f}'} \mathcal{D} \vec{Y}'  \rho( \vec{Y}_{ i}, \vec{Y}_{ i}'; t_i)
\end{eqnarray}
adopted to keep long expressions compact
where  the integrals over $\vec{Y}_i$ and $\vec{Y}_i'$ trace over all configurations of the field $\vec{Y}$ at the initial time.
To capture the averaged effect of the reservoir on the matter we trace over it's final variables
leading to the reservoir-reduced influence functional, $\mathcal{F}_X[ \vec{P}, \vec{P}']$

\begin{eqnarray}
\label{RRDM}
Tr_{X} \{  \rho( \vec{P}_{ f}, \vec{P}_{ f}',\vec{X}_{\nu f}, \vec{X}_{\nu f}'; t) \} 
=
  \int^{\vec{P}_f, \vec{P}_f'}_{CTP} \mathcal{D} \vec{P} \
  \exp \{ i(S_M[\vec{P}]-S_M[\vec{P}'] +S^{PF}_{int}[A^\mu, \vec{P}]-S^{PF}_{int}[A^{\mu'}, \vec{P}']) \} 
  \nonumber \\
  \times
\mathcal{F}_X[ \vec{P}, \vec{P}'].
\end{eqnarray}

The reservoir-reduced  influence action is given by
%\int_V d^3x \int_{t_i}^{t_f} dt

\begin{eqnarray}
\label{ }
-i \ln \mathcal{F}_X[ \vec{P}, \vec{P}'] \equiv S^X_{IF}[\vec{P}, \vec{P}'] =   \int_V d^4 x \int_V d^4 x' 
\left[ P^{i-}(t, \vec{x}) \frak{G}^{ret}(x,x') P^{+}_i(t', \vec{x}) + \frac{i}{4}  P^{i-}(t, \vec{x}) \frak{G}^{H}(x, x') P^{-}_i(t', \vec{x}) \right]
\end{eqnarray}
where $\frak{G}^{ret}(x,x')$ and $\frak{G}^{H}(x,x')$ are the retarded and Hadamard Green's functions for the reservoir, respectively, 
and the superscripts $+$ and $-$  denote semi-sum  and  difference variables
defined by $P^+=(P+P')/2$ and $P^-=P-P'$ respectively. Here the unprimed and primed quantities denote 
the forward and backward histories in the closed-time-path
(or Schwinger-Keldysh) sense \cite{Schwinger61, Keldysh65}. 

The reservoir kernels take the explicit form

\begin{equation}
\label{ }
\frak{G}^{ret}(x,x') =  \int_0^\infty d \nu \ (g^2(\nu)/ \nu I(\nu) ) \sin \nu(t-t') \theta(t-t') \delta^3(\vec{x}- \vec{x}')  \ \ \ \ \ \vec{x} \in V
\end{equation}
and 
\begin{equation}
\label{ }
 \frak{G}^{H}(x, x') =  \int_0^\infty d \nu \ (g^2(\nu)/ \nu I(\nu) ) \coth (\beta_X \nu/2) \cos \nu(t-t')\delta^3(\vec{x}- \vec{x}') \ \ \ \ \ \vec{x} \in V
\end{equation}
where we have assumed that the initial state of the reservoir is thermal with inverse temperature $\beta_X$. The matter is not spatially correlated because the reservoir oscillators at one position  do not interact with their neighbors.   
The retarded Green's function represents the response of the reservoir to an external field and so corresponds to its susceptibility.  The Hadamard function, which can be derived from the symmetric two-point function of the reservoir's variables, quantifies reservoir fluctuations. This can be seen by noting that when the time arguments are coincident the Hadamard function becomes

\begin{equation}
\label{ }
\frak{G}^{H}(t, t) = \int_0^\infty d \nu \ (g^2(\nu)/ \nu I(\nu) ) \left< \Delta X_\nu^2 (t) \right>
\end{equation}
where $\frak{G}^{H}(x,x') = \frak{G}^{H}(t,t') \delta^3(\vec{x} - \vec{x}')$ and $X_\nu (t)$ is the displacement of the $\nu$th reservoir oscillator at a point.

  By the assumption that the reservoir is initially in thermal equilibrium these kernels satisfy the fluctuation-dissipation relation. Taking the Fourier transform of the time dependent component of each kernel we find

\begin{equation}
\label{ }
\frak{G}^{ret}(\omega, \vec{x}, \vec{x}') =  \int_0^\infty d \nu \ (g^2(\nu)/ \nu I(\nu) ) \bigg[ \frac{\nu}{\nu^2 - \omega^2} + i \frac{\pi}{2}\bigg( \delta(\omega - \nu) - \delta (\omega + \nu) \bigg) \bigg] \delta^3(\vec{x}- \vec{x}') 
\end{equation}
and 
\begin{equation}
\label{ }
 \frak{G}^{H}( \omega,  \vec{x}, \vec{x}' ) = \pi  \int_0^\infty d \nu \ (g^2(\nu)/ \nu I(\nu) ) \coth (\beta_X \nu/2) 
\bigg[  \delta(\omega - \nu) + \delta (\omega + \nu) \bigg]
 \delta^3(\vec{x}- \vec{x}') 
 \end{equation}
 noting that the imaginary part of the reservoir susceptibility $\frak{G}^{ret}(\omega, \vec{x}, \vec{x}')$, which describes the absorption of energy by the reservoir, is balanced by the reservoir's fluctuations
 
 \begin{equation}
\label{FDR_RM}
 \frak{G}^{H}( \omega,  \vec{x}, \vec{x}' )  = 2  \coth (\beta_X \omega /2) \text{Im}[ \frak{G}^{ret}(\omega,  \vec{x}, \vec{x}') ].
\end{equation}

To see explicitly how reservoir fluctuations influence the matter we appeal to a
Gaussian path integral identity first suggested by Feynman and Vernon
\cite{FeyVer}. Note the complex modulus of the influence functional
can be written as

\begin{equation}
\label{ }
| \mathcal{F}_X[\vec{P}, \vec{P}'] |= \int \mathcal{D} \vec{\xi}_X \mathcal{P}[\vec{\xi}_X] \exp \bigg\{i \int_V d^4 x  \ \vec{\xi}_X(x) \cdot \vec{P}^{-}(x) \bigg\}
\end{equation}
where $\mathcal{P}[\vec{\xi}_X]$ takes the  form (with a normalization constant) 
\begin{equation}
\label{ }
\mathcal{P}[\vec{\xi}_X]=\exp \bigg\{ - \int_V d^4 x   \int_V d^4 x' \ \delta_{jk} \ \xi_X^j( x ) \frak{G}^{-1}_H( x, x') \xi^k_{X}( x') \bigg\}
\end{equation}
allowing the reservoir-reduced influence functional to be expressed as
\begin{equation}
\label{ }
\mathcal{F}_X[\vec{P}, \vec{P}'] = \int \mathcal{D} \vec{\xi}_X \mathcal{P}[\vec{\xi}_X] \exp \bigg\{ i \int_V d^4 x \ P^{-}_i(x) [ \xi^i_X(x) + \int_V d^4 x' \frak{G}^{ret}(x, x') P^{i+}(x') ]  \bigg\}.
\end{equation}
In this process we have replaced the kernel describing quantum and thermal fluctuations
in the reservoir with the classical variable $\vec{\xi}_X$ which drives the field in the same manner as an external stochastic force. 
To retrieve information
about the reservoir's fluctuations it is necessary to integrate over the
functional distribution  $\mathcal{P}[\vec{\xi}_X]$ which is positive definite because
%in this sense the field $\xi_j$ has no meaning independent of the functional distribution $P[\vec{\xi}]$.
the kernel $\frak{G}^{-1}_H$ is symmetric and positive.
Therefore
we interpret the field $\vec{\xi}_X$ as a classical stochastic force or noise, driving the matter
with probability distribution described by $\mathcal{P}[\vec{\xi}_X]$. Due to
the Gaussianity of $\mathcal{P}[\vec{\xi}_X]$ all of its moments are specified
by the mean $\left< \xi_X^j \right>_{\xi_X}=0$ and the variance $\left<
\{ \xi_X^j(x), \xi_X^k(x') \} \right>_{\xi_X}= \delta^{jk} \frak{G}_H(x, x')$ where $\left< ... \right>_{\xi_X}=\int
\mathcal{D} \vec{\xi}_X \mathcal{P}[\vec{\xi}_X] (...)$.

Now that we have coarse-grained over the reservoir we can derive the semi-classical equation of motion for the matter field,
ignoring for now its interaction with the field, to see the reservoir's averaged effect on the matter. A saddle point approximation of 
(\ref{RRDM}) after the fluctuation kernel has been replaced by a stochastic forcing term gives 

\begin{equation}
\label{MEOM}
\ddot{P}^k + \omega^2 P^k - \int_{t_i}^{t_f} dt' \frak{G}^{ret}(t,t') P^k(t') = \xi^k_X(t).
\end{equation}
The effect of the reservoir is now manifest; first, the addition of the  integral kernel in the equation of motion
leads to dissipation in the free oscillations of the matter field, and second, the fluctuations of the reservoir drive the matter field through the source
 term on the right. For the specific case of Ohmic reservoir spectral density, that is $g^2(\nu)/ \nu I(\nu) = 2 \gamma \nu /\pi$, we find

\begin{equation}
\label{SCEOM_MF}
\ddot{P}^k + \tilde{\omega}^2 P^k + \gamma \dot{P}^k(t') = \xi^k_X(t)
\end{equation}
where the integral kernel leads to a dissipative term and a frequency renormalization $\tilde{\omega}^2 = \omega^2 - 2 \gamma \delta(0)$
and the fluctuation kernel defining the distribution of $\xi^k_X(t)$ becomes

\begin{equation}
\label{ }
 \frak{G}^{H}(t,t') =  \frac{ 2 \gamma}{\pi}  \int_0^\infty d \nu \  \nu \coth (\beta_X \nu/2) \cos \nu(t-t')\delta^3(\vec{x}- \vec{x}') \ \ \ \ \ \vec{x} \in V.
\end{equation}

%%%%%%%%%%%%%%%%%%%%%%%%%%%%%%%%%%%%%%%%%%%%%%%%%%%%%%%%%%
\subsection{Medium-Reduced Density Matrix}
%%%%%%%%%%%%%%%%%%%%%%%%%%%%%%%%%%%%%%%%%%%%%%%%%%%%%%%%%%

%We have now coarse-grained over the reservoir. To complete the medium influenced description of the field 
%we need to continue this process by tracing over the matter degrees of freedom in the total density matrix.  
%

Consider now the time evolution of the density matrix elements describing the
combined field + medium system, where the reservoir has already been integrated out, expressed as a product
of path integrals

\begin{eqnarray}
\label{ }
\rho(A^\mu_f, A^{\mu'}_f, \vec{P}_{f}, \vec{P}_{f}'; t)
=
  \int^{A^\mu_f, A^{\mu'}_f}_{CTP} \mathcal{D}A^\mu 
   \
\exp \{ i(S_E[A^\mu]- S_E[A^{\mu'}] ) \}
  \nonumber
\\
\times
 \int \mathcal{D} \vec{\xi}_X \mathcal{P}[\vec{\xi}_X] 
 \int^{\vec{P}_f, \vec{P}_f'}_{CTP} \mathcal{D} \vec{P}  
\exp \bigg\{ i \bigg[
\tilde{S}_M[\vec{P}]-\tilde{S}_M[\vec{P}'] +S^{PF}_{int}[A^\mu, \vec{P}]-S^{PF}_{int}[A^{\mu'}, \vec{P}']
+
 \int_V d^4 x  \ \vec{\xi}_X \cdot \vec{P}^- 
\bigg]
\bigg\} .
\end{eqnarray}
The action $\tilde{S}_M$ contains the dissipation term from the reservoir-reduced influence action

\begin{equation}
\label{ }
\tilde{S}_M[\vec{P}] =S_M[\vec{P}] + \frac{1}{2}\int_V d^4 x  \int_V d^4 x' \ P^i(x) \frak{G}^{ret}(x, x')  P_i(x').
\end{equation}
Upon tracing  the density matrix of the combined field + medium system 
over the matter variables  we obtain a reduced density matrix $\rho_r=Tr_{\vec{P}} \rho$ that accounts for the averaged effect the medium has on the field. 
For linear coupling
between the medium and the field and an initially Gaussian matter
state the path integrals 
%for the dielectric 
can be evaluated exactly
yielding the medium-reduced density matrix
\cite{FN1}.
%{We should point out here that when the medium and the field interact very strongly the assumption that the state is initially factorizable may not be a good assumption.  We adopt such a state here for ease of computation.  For more general situations one can introduce a preparation function to describe the initial state,  but as shown in \cite{Romero97}
%the key features are not affected too strongly. We will return to this issue later.}.
The result of tracing over matter variables gives

\begin{eqnarray}
\label{mrdm}
\rho_r(A^\mu_f, A^{\mu'}_f; t)= 
  \int^{A^\mu_f A^{\mu'}_f}_{CTP} \mathcal{D}A^\mu 
\exp \{ i(S_E[A^\mu]- S_E[A^{\mu'}]  \} 
\mathcal{F}_M[A^\mu, A^{\mu'}]
\end{eqnarray}
where the medium-reduced influence functional
$\mathcal{F}_M[A^\mu, A^{\mu'}] $ accounts for the averaged
effect of the medium on the field. The influence action $S^M_{IF} $, which relates to the influence functional as 
$ \mathcal{F}_M =  \int \mathcal{D} \vec{\xi}_X \mathcal{P}[\vec{\xi}_X] 
\exp\{ i S^M_{IF} \}$, is given by

\begin{equation}
\label{ }
S^M_{IF}[A^\mu, A^{\mu'}] = \int_V d^4x  \int_V d^4 x' \  E^{-}_i(x) \bigg\{  \tilde{g}^{ret}(x,x') \bigg[ \Omega_P^2 E^{i+}(x') + \Omega_P \xi^i_X(x')\bigg]  +\frac{i}{4} \Omega_P^2 \tilde{g}^{H}(x,x') E^{i-}(x') \bigg\}.
\end{equation}
The kernels $\tilde{g}^{ret}(x,x')$ and $\tilde{g}^H(x,x')$ are the matter's retarded and Hadamard functions, respectively. In this case the retarded Green's function does not follow directly from the commutator of the matter's polarization operators because the semi-classical equation of motion for the reservoir-influenced polarization field is not self-adjoint. It can be solved for classically or by considering the commutator of the polarization  operator with its adjoint with respect to (\ref{MEOM}).
Because the matter + reservoir system is linear the Heisenberg equation of motion for the polarization operators is the same as the semi-classical equation (\ref{SCEOM_MF}). The symmetric two-point function of the homogeneous solution of this equation subject to the appropriate boundary conditions provides $\tilde{g}^H(x,x')$.
Because the oscillators comprising the matter do
not interact and the medium is assumed to be isotropic, the
polarization fluctuations within the medium are not spatially
correlated i.e. $\tilde{g}(x,x')=\tilde{g}(t,t') 
\delta^3(\vec{x}-\vec{x}')$. Under these conditions the influence
action simplifies greatly

\begin{equation}
\label{ }
S^M_{IF}[A^\mu, A^{\mu'}] = \int_V d^4 x \int_{t_i}^{t_f} dt'  \ E^{-}_j(t,\vec{x}) \bigg\{ \tilde{g}_{ret}(t,t') \bigg[ \Omega_P^2 E^{j+}(t',\vec{x}) + \Omega_P \xi^j_X(t, \vec{x}) \bigg] +\frac{i}{4} \Omega_P^2 \tilde{g}_{H}(t,t') E^{j-}(t',\vec{x}) \bigg\}.
\end{equation}
%It is instructive to write the solution for $\tilde{g}_{ret}(t,t')$ in frequency space (where we have taken the infinite time limit in order to take the Fourier transform)
%
%\begin{equation}
%\label{ }
%\tilde{g}_{ret}(\omega') = - \frac{1}{ {\omega'}^2 - \omega^2 +  \frak{G}^{ret}(\omega')}
%\end{equation}
%
where the explicit form for $\tilde{g}_{ret}(t,t')$ and $\tilde{g}_H(t,t')$ are given below for Ohmic spectral density of the reservoir

\begin{equation}
\label{dis}
\tilde{g}_{ret}(t,t')= \frac{1}{\overline{\omega}}e^{- \gamma/2(t-t') }\sin \overline{\omega} (t-t') \theta(t-t')
\end{equation}

\begin{equation}
\label{fluc}
\tilde{g}_{H}(t,t')= \frac{1}{ \tilde{\omega} \overline{\omega}^2 }e^{- \gamma/2(t+t' -2 t_i)} \coth(\beta_P \omega/2 ) 
\bigg[
\tilde{\omega}^2 \cos \overline{\omega} (t-t') 
-
\frac{ \gamma^2}{4}  \cos \overline{\omega} (t + t' - 2t_i ) 
+
\frac{ \gamma}{2}  \overline{\omega} \sin \overline{\omega} (t + t' - 2t_i) 
\bigg]
\end{equation}
where $\overline{\omega} = \sqrt{ \tilde{\omega}^2- \gamma^2/4}$ and $t_i$ is the initial time at which the reservoir and matter begin to interact. Note in particular that the Hadamard function is not stationary and thus describes a nonequilibrium process. It can easily be verified that when the matter-reservoir coupling goes to zero 
these kernels give back the expected free forms. 

As with the case of coarse-graining over the reservoir described in the previous section we see that
tracing over the matter degrees of
freedom leads to two terms that are nonlocal in time: 
 the term
containing the retarded Green's function characterizes the response
of the dielectric to the field ($\tilde{g}_{ret}$ plays the role of the
susceptibility), and the one containing the Hadamard function
quantifies the effects that quantum and thermal fluctuations
($\beta_P$ being the matter's inverse temperature) of the
polarization within the medium has upon the field. 

As with the case of the reservoir we can represent the fluctuation 
kernel in terms of a stochastic force. 
This allows us to write the complex modulus of the influence functional as

\begin{equation}
\label{ }
| \mathcal{F}_M[A^\mu, A^{\mu'}] |= \int \mathcal{D} \vec{\xi}_M \mathcal{P}[\vec{\xi}_M] \exp \bigg\{i \Omega_P \int_V d^4 x \ \xi_M^i(x) E^{-}_i(x) \bigg\}
\end{equation}
where $\mathcal{P}[\vec{\xi}_M]$ takes the  following form modulo a normalization constant 
\begin{equation}
\label{ }
\mathcal{P}[\vec{\xi}_M]=\exp \bigg\{ - \int_V d^4 x  \int_V d^4 x'  \ \delta_{ij} \xi^i_M(x) \tilde{g}^{-1}_H( x,x') \xi^j_M(x') \bigg\}
\end{equation}
which defines $\left<\{ \xi_M^j(x), \xi_M^k(x') \} \right>_{\xi_M}= \delta^{jk}  \tilde{g}_H(t,t')
\delta^3(\vec{x}-\vec{x}')$.

Using $\mathcal{P}[\vec{\xi}_M]$ the medium-reduced influence functional becomes
\begin{eqnarray}
\label{ }
\mathcal{F}_M[A^\mu, A^{\mu'}]=    \int \mathcal{D} \vec{\xi}_X \mathcal{P}[\vec{\xi}_X]  \int \mathcal{D} \vec{\xi}_M \mathcal{P}[\vec{\xi}_M] \exp \bigg\{ i  \Omega_P \int_V d^4 x \ E^{-}_i(x) 
\bigg[
  \xi^i_M(x)
  \nonumber \\
   + \int_{t_i}^{t_f} dt' \ \tilde{g}^{ret}(t,t')( \Omega_P E^{i+}(t, \vec{x}) + \xi^i_X(t, \vec{x}))
 \bigg]  \bigg\}.
\end{eqnarray}

Now we turn to the stochastic effective action $S_\xi[A^\mu, A^{\mu'}]$ for the field
which is the sum of the free-field action and the influence action
after the terms quantifying the fluctuations of the
medium  have been replaced with stochastic forces. As we
work with the photon field in the path integral a gauge-fixing
prescription must be adopted in order to prevent summing over gauge equivalent orbits.
 The Fadeev-Popov trick could be
employed but the ghost fields introduced do not couple to the
electromagnetic field in flat space and only contribute an overall
multiplicative constant. Furthermore, the currents in our microscopic
model which couple to the photon field are conserved. So we are free
to choose any gauge we wish to evaluate the path integral \cite{BD}. As a
result the Green's functions which appear will be gauge dependent.
However, as a consequence of current conservation any gauge choice
will give equally valid descriptions of physical processes. With this freedom we
can express the stochastic effective action for the field, after
choosing the temporal gauge ($A^0=0$). At this level we can make a comparison 
with MQED by allowing $t_f$ and $-t_i$ to go to infinity, assuming the 
field evolves to a steady-state and taking the Fourier
transform to find 

\begin{eqnarray}
\label{ }
S_{\xi}[A^\mu, A^{\mu'}]=\frac{1}{2\pi} \int_{-\infty}^\infty d {\omega'} \int d^3x \bigg[-
(\nabla \times \vec{A}^-(-{\omega'},\vec{x}))\cdot
(\nabla \times \vec{A}^+({\omega'},\vec{x}))
\nonumber \\
+ \vec{A}^{-}(-{\omega'},\vec{x}) \cdot \vec{A}^{+}({\omega'},\vec{x})
{\omega'}^2(1+\Omega_P^2 \tilde{g}_{ret}({\omega'}, \vec{x})) + \Omega_P \vec{E}^{-}(-{\omega'},\vec{x}) \cdot
\tilde{g}_{ret} ({\omega'}, \vec{x}) \vec{\xi}_X({\omega'},\vec{x}) \bigg]
\end{eqnarray}
where in the infinite time limit the matter's stochastic force is exponentially suppressed $\vec{\xi}_M \rightarrow 0$.
Note that we've added the explicit
position dependence of the medium's retarded Green's function because its support 
is restricted to the volume containing the dielectric.

The stochastic semi-classical equations of motion for the field can
be derived  from the saddle point condition for the medium-reduced density
(\ref{mrdm})
matrix which gives

\begin{equation}
\label{SCEM} \nabla \times \nabla \times \vec{A}({\omega'}, \vec{x}) -
{\omega'}^2(1+ \Omega_P^2 \tilde{g}_{ret}({\omega'}, \vec{x})) \vec{A}({\omega'}, \vec{x}) =  i \Omega_P {\omega'}
 \tilde{g}_{ret} ({\omega'}, \vec{x}) \vec{\xi}_X({\omega'},\vec{x}) .
\end{equation}
In this form the role of the coarse-grained medium is evident.  By comparing with the
macroscopic Maxwell's equations  we identify the
permittivity  as $\varepsilon({\omega'},
\vec{x})=1+\Omega_P^2 \tilde{g}_{ret}({\omega'}, \vec{x})$ and observe that the fluctuations of the medium drive
the field through the stochastic force(s) $\vec{\xi}_X(x)$ ($\vec{\xi}_M(x)$ drives the field in addition to the reservoir when the system is not in a steady-state). For Ohmic spectral density of the reservoir the frequency dependent permittivity corresponds with the Lorentz oscillator model

\begin{equation}
\label{ }
\varepsilon({\omega'}, \vec{x}) = 1 - \frac{\Omega_P^2}{\omega' (\omega' + i \gamma) - \tilde{\omega}^2} \ \ \ \ \  \vec{x} \in V.
\end{equation}
When the restoring force of the matter vanishes we obtain the Drude model and the plasma model when $\gamma \rightarrow 0$  in addition. In this form it's clear that the matter-field coupling  $\Omega_P$ can be interpreted as the medium's plasma frequency. 

One can see from equation (\ref{SCEM}) a striking similarity with
MQED. Due to the linearity of our theory this is not surprising.
Indeed if one were to proceed from this point treating the field semi-classically
and choosing all space to be filled with a dielectric material, albeit in vacuum this dielectric
is fictitious, one would exactly reproduce the predictions of MQED using (\ref{SCEM}) and 
choosing the dielectric to be in a thermal state.
However, it is important to note that in this case
the stochastic field $\xi^j_X$ represents the fluctuations of the
medium only and does not include the intrinsic fluctuations of the field.
After the next level of coarse-graining described in the following
section the intrinsic
quantum fluctuations of the field will enter which is different from
the induced fluctuations from interaction with the dielectric medium.

 %%%%%%%%%%%%%%%%%%%%%%%%%%%%%%%
\section{Field-Reduced Density Matrix}
 %%%%%%%%%%%%%%%%%%%%%%%%%%%%%%%

In the last section we showed how the medium influences the field, and for the particular 
case of local fluctuations we found that the stochastic semi-classical action for the field 
takes the same form as the noninteracting field (with frequency dependent velocity) driven 
by an external current (stochastic force).

A detailed description of the field's microstate is unimportant in the
description of the atom-surface force. %and in principle are beyond our measuring capabilities. 
By coarse-graining over the medium-influenced field we can incorporate the averaged effect
of the field on the atom's trajectory without a specific knowledge of
the final field state, leading to the reduced density matrix for
the atom

\begin{eqnarray}
\label{ }
\rho_r(z_f,z_f', \vec{Q}_f, \vec{Q}_f'; t_f)= 
 \int^{\vec{z}_f, \vec{z}_f'}_{CTP} \mathcal{D} \vec{z}  
 \int^{\vec{Q}_f, \vec{Q}_f'}_{CTP} \mathcal{D} \vec{Q}  
  \exp \{ i (S_Z[\vec{z}]  -S_Z[\vec{z}'] +S_Q[\vec{Q}] -S_Q[\vec{Q}'] \}
    \nonumber \\  
    \times
 \int  \mathcal{D} A^\mu_f  
  \int^{A^\mu_f, A^{\mu}_f}_{CTP} \mathcal{D}A^\mu \
\exp\{ S^{AF}_{int}[ \vec{z}, \vec{Q},  A^\mu]-S^{AF}_{int}[ \vec{z}' , \vec{Q}' ,  A^{\mu'}] ) \} 
\mathcal{F}_M[A^\mu, A^{\mu'}]
\end{eqnarray}
where the first integral in the second line traces over the final field configurations.

For the assumed initially factorized state the path integrals
over the field can be evaluated exactly if the initial state is
Gaussian yielding the field-reduced influence functional,
$\mathcal{F}_\xi[J^\mu, J^{\mu'}, \xi] = \int \mathcal{D} \vec{\xi}_X \mathcal{P}[\vec{\xi}_X]  \int \mathcal{D} \vec{\xi}_M \mathcal{P}[\vec{\xi}_M] \exp\{ iS^E_{IF}[J^\mu, J^{\mu'},
\xi] \} $ expressed in
terms of the  influence action, $S^E_{IF}$, given by

\begin{eqnarray}
\label{SEIF}
S^E_{IF}[J^\mu, J^{\mu'}, \xi^j]=
 \int d^4 x \int d^4 x' J^{\mu-}({x}) \bigg[ \tilde{D}^{ret}_{\mu \nu}({x}, {x}') \big(J^{\nu+}({x}')-\kappa^\nu_i \xi^i({x}')\big)
  +\frac{i}{4} \tilde{D}^{H}_{\mu \nu}({x},{x}') J^{\nu-}(\vec{x}') \bigg]
\end{eqnarray}
where $\vec{\xi}(x)  \equiv     \vec{\xi}_M(x) + \int_{t_i}^{t_f} dt' \ \tilde{g}_{ret} (t,t') \vec{\xi}_X(t',\vec{x}) $.
%For later ease we give the influence action in Fourier space where @@ the existence of a stationary state at late times @@ is implicitly assumed
%
%\begin{eqnarray}
%\label{SEIF2}
%S^E_{IF}[J^\mu, J^{\mu'}, \xi^j]=
 %\frac{1}{2\pi}\int d\omega d^3 x d^3 x' J^{\mu-}(-\omega, \vec{x}) \bigg[ \tilde{D}^{ret}_{\mu \nu}(\omega, \vec{x}, \vec{x}') \big(J^{\nu+}(\omega,\vec{x}')-\kappa^\nu_i \xi^i(\omega, \vec{x}')\big)
% \nonumber \\
%  +\frac{i}{4} \tilde{D}^{H}_{\mu \nu}(\omega, \vec{x}, \vec{x}') J^{\nu-}(\omega,\vec{x}') \bigg].
%\end{eqnarray}
%
The current density $J^\mu$ in (\ref{SEIF}) comes from the atom-field interaction and takes the explicit form
\begin{equation}
\label{ }
J^\mu(x)=-q \int d\lambda \ Q^i(\lambda) \kappa_i^\mu \delta^4(x^\alpha-z^\alpha(\lambda))
\end{equation}
where the derivative operator $\kappa^\mu_i=-\partial_0 \eta^\mu_i+\partial_i \eta^\mu_0$ yields the electric field when contracted with the vector potential $E^i =\kappa_\mu^i A^\mu$ and also enforces current conservation $\partial_\mu \kappa^\mu_i f(x)=(-\partial_0 \partial_i+\partial_i \partial_0)f(x)=0 $.
The integral kernels $\tilde{D}^{ret}_{\mu \nu}$ and $\tilde{D}^{H}_{\mu \nu}$ are
the retarded  Green's and Hadamard function for the medium-altered
electromagnetic field which result from solving the semi-classical
equations of motion (for the retarded Green's function sourced by a delta function). 
The retarded Green's function describes the classical electrodynamical propagation
of the field in the presence of the dielectric material, and the Hadamard function describes
the field's intrinsic quantum fluctuations.
%As we noted in the
%previous section the semi-classical equations describe a driven field in
%the presence of a medium described by a frequency- and spatially-
%dependent permittivity and so the field must satisfy the boundary conditions
%associated with the geometry of the dielectric.

In the temporal gauge the semi-classical equation of motion for the field's retarded Green's
function  in the presence of a dielectric medium satisfies 

\begin{equation}
\label{GFeqn}
 \epsilon^{ab}_{\ \ \ i}   \epsilon^{mn}_{\ \ \
\ b} \partial_a \partial_m  \tilde{D}^{ret}_{nk}(x,x')  + \frac{\partial^2}{ \partial t^2 }   \tilde{D}^{ret}_{ik}({x}, {x}')  +  \frac{\partial}{ \partial t} \int_{t_{in}}^{t_f} dt_2 \ \tilde{g}_{ret}(t,t_2;\vec{x})
\frac{\partial}{ \partial t_2}
\tilde{D}^{ret}_{ik}(t_2 ,\vec{x},  x') =  \delta_{ik} \delta^4({x}-{x}')
\end{equation}
%
%and in frequency representation as
%
%\begin{equation}
%\label{GFeqn_freq}
% \epsilon^{ab}_{\ \ \ i}   \epsilon^{mn}_{\ \ \
%\ b} \partial'_a \partial'_m  \tilde{D}^{ret}_{nk}(\omega, \vec{x},  \vec{x}') - \omega^2 \epsilon(\omega, \vec{x})  \tilde{D}^{ret}_{ik}(\omega, \vec{x},  \vec{x}') =  \delta_{ik} \delta^3(\vec{x}-\vec{x}')
%\end{equation}
where $\epsilon_{abc}$ is the Levi-Civita symbol (Roman indices
refer to spatial components).
The solution to (\ref{GFeqn}) gives the particular solution, $A^P_j$, to the semiclassical equation of motion (\ref{SCEM}) for the electromagnetic field

\begin{equation}
\label{ }
A^P_j(x)= -   \int d^4 x'   \tilde{D}^{ret}_{jk}(x, x') \frac{ \partial }{ \partial t'}  \xi^k( x').
\end{equation}

As was noted previously the Heisenberg equations of motion for the field operators take the same form as the semi-classical equation of motion because of the linear coupling

\begin{equation}
\label{HEOM}
 \epsilon^{ab}_{\ \ \ i}   \epsilon^{mn}_{\ \ \
\ b} \partial_a \partial_m  \hat{A}_n(x)  + \frac{\partial^2}{ \partial t^2 }   \hat{A}_i(x)  +  \frac{\partial}{ \partial t} \int_{t_{in}}^{t_f} dt_2 \ \tilde{g}_{ret}(t,t_2;\vec{x})
\frac{\partial}{ \partial t_2}
 \hat{A}_i(t_2, \vec{x})   =  \frac{ \partial }{ \partial t}  \xi_i( x)
 \end{equation}
 therefore we can use the homogeneous solution to (\ref{HEOM}) to construct the symmetric two-point function of the field operators to give the Hadamard function
 
 \begin{equation}
\label{ }
 \tilde{D}_{H}^{i j}(x, x') = \left<  \{ \hat{A}^i_o(x),  \hat{A}^j_o(x') \} \right>. 
\end{equation}

Finally, the field-reduced density matrix takes the form

\begin{eqnarray}
\label{ }
\rho_r(z_f,z_f', \vec{Q}_f, \vec{Q}_f'; t_f)=  
 \int^{\vec{z}_f, \vec{z}_f'}_{CTP} \mathcal{D} \vec{z}  
 \int^{\vec{Q}_f, \vec{Q}_f'}_{CTP} \mathcal{D} \vec{Q} 
 \ e^{i (S_Z[\vec{z}] +S_Q[\vec{Q}]-S_Z[\vec{z}'] -S_Q[\vec{Q}'])} \mathcal{F}_\xi[J^\mu, J^{\mu'}, \xi].
\end{eqnarray}

 %%%%%%%%%%%%%%%%%%%%%%%%%%%%%%%
\section{Atom's internal dof (Q)-Reduced Density Matrix}
 %%%%%%%%%%%%%%%%%%%%%%%%%%%%%%%

At this point we have the density matrix that describes the dynamics
of the atom's trajectory and its internal degrees of freedom under
the influence of the medium-altered field. As with the previous parties it is only the
averaged effect and not the microscopic details of the atom's
internal degree of freedom which we need for the description of the
force. 

As our last tier of coarse-graining we
trace over $\vec{Q}_f$ the oscillator's internal dof to obtain the
oscillator-reduced density matrix which characterizes the
dynamics of the atom's trajectory determined by its interaction with
all remaining parties
\begin{equation}
\label{rhoZ}
\rho_Z(z_f,z_f'; t_f)=  
 \int^{\vec{z}_f, \vec{z}_f'}_{CTP} \mathcal{D} \vec{z} \ e^{i (S_Z[\vec{z}] -S_Z[\vec{z}'] )} \mathcal{F}_Z[\vec{z}, {\vec{z}}'].
\end{equation}
The environmental influences are now packaged in the
oscillator-reduced influence functional $\mathcal{F}_Z[\vec{z},
\vec{z}']$ with their back-action accounted for in a self-consistent
manner. The coarse-grainings are summarized in Table \ref{table:CG}
\begin{table}[ht] 
\caption{Summary of Coarse-Graining and Detailed Physics Remaining at Each Tier} % title of Table 
\centering % used for centering table 
\begin{tabular}{ccccc} % centered columns (4 columns) 
\hline\hline %inserts double horizontal lines 
\ \ \ Tier \ \ \ & \ \ \ Coarse-Graining \ \ \ &  \ \ \ Influence Functional \ \ \ & \ \ \ Detailed Physics \ \ \  & \ \ \ Remaining Variables \ \ \  \\ [1.5ex] % inserts table %heading \hline % inserts single horizontal line 1 & 50 & 837 & 970\\ % inserting body of the table
0 &                &                                 & \text{all parties}                     & $\vec{z}, \vec{Q}, A^\mu, \vec{P}, \vec{X}_\nu$    \\
1 & $Tr_X$ & $\mathcal{F}_X$   & \text{atom + field + matter}  & $\vec{z}, \vec{Q}, A^\mu, \vec{P}$    \\
2 & $Tr_P$ & $\mathcal{F}_M$  & \text{atom + field}                  & $\vec{z}, \vec{Q}, A^\mu$  \\
3 & $Tr_A$ & $\mathcal{F}_\xi$ & \text{atom}                              & $\vec{z}, \vec{Q}$  \\ 
4 & $Tr_Q$ & $\mathcal{F}_Z$  & \text{atom's motion}              & $\vec{z}$ \\ [1ex]
% [1ex] adds vertical space 
\hline %inserts single line 
\end{tabular} 
\label{table:CG} % is used to refer this table in the text 
\end{table}
at each tier the influence of all lower tiers on the remaining degrees of freedom is packaged in the influence functional. 

We proceed by evaluating $\mathcal{F}_Z[\vec{z}, \vec{z}']$
perturbatively to lowest order in the atom-field coupling. In the fully dynamical case an exact calculation 
is not possible as backaction of the field on the dynamics of the atom's dipole moment
enters as a third time derivative and includes multiple reflections between the dielectric medium and the atom. 
At leading order in powers of the atom-field coupling, an expression for the force can be found which neglects
the radiation reaction to the atom's dipole moment and the effects of multiple reflections.

To ease the notational burden we define

\begin{equation}
\label{ }
\left<...\right>_o = \int_{-\infty}^\infty d\vec{Q}_f  
 \int^{\vec{Q}_f, \vec{Q}_f}_{CTP} \mathcal{D} \vec{Q} 
  \ e^{i (S_Q[\vec{Q}]-S_Q[\vec{Q}'])} (...)
\end{equation}
which represents the noninteracting time-dependent expectation
value with respect to the oscillator's initial state. With this
simplification the oscillator-reduced influence functional can be
compactly expressed in terms of quantum and stochastic expectation
values
\begin{equation}
\label{IFZ}
e^{iS_{IF}[\vec{z}, \vec{z}']}  \stackrel{def}{=}  \mathcal{F}_Z[\vec{z}, \vec{z}']= \bigg< \bigg<  \bigg< e^{i S^E_{IF}} \bigg>_o \bigg>_{\xi_M} \bigg>_{\xi_X}
\end{equation}
which introduces the influence action, $S_{IF}$.

A saddle point approximation of (\ref{rhoZ}) gives the semi-classical equation of motion for the trajectory where the saddle point is determined by the equation

\begin{equation}
\label{ }
\frac{\delta}{ \delta z^{k-}(\tau)} S_{CGEA}[\vec{z},\vec{z}']\bigg|_{z^{k-}=0}=0 \Rightarrow M\ddot{z}_k(\tau)+\partial_k V[\vec{z}(\tau) ]=f_k(\tau),
\end{equation}
the coarse-grained effective action, $S_{CGEA}[\vec{z},\vec{z}']$, is defined as
$S_{Z}[\vec{z}]-S_{Z}[\vec{z}']+S_{IF}[\vec{z},\vec{z}']$,
and the influence force, $f_k(\tau)$, is given by

\begin{equation}
\label{ }
\frac{\delta}{ \delta z^{k-}(\tau)} S_{IF}[\vec{z},\vec{z}']\bigg|_{z^{k-}=0}=f_k(\tau).
\end{equation}

We now wish to evaluate the influence force perturbatively for an expansion in terms of small atom-field coupling. We begin by expanding both sides of (\ref{IFZ}). 
Because $S^E_{IF}$ contains terms linear and quadratic in the atom-field
coupling we find  to $\mathcal{O}(q^3)$,

\begin{equation}
\label{ } \mathcal{F}_Z[\vec{z}, \vec{z}'] = 1+iS_{IF}+ \mathcal{O}(q^3)  \approx 1+ i \left< \left<
S^E_{IF} \right>_o \right>_\xi -\frac{1}{2} \left< \left< ( S^E_{IF}
)^2 \right>_o \right>_\xi+\mathcal{O}(q^3) 
\end{equation}
where the subscript $\xi$ in $ \left< ...\right>_\xi$ represents that both the $\xi_X$ and $\xi_M$ stochastic expectation values have been taken. 
On the right hand side there are two contributions:  the term linear in $S^E_{IF}$ reduces to

\begin{eqnarray}
\label{ }
\left< \left<  S^E_{IF} \right>_o \right>_\xi=
  \int d^4 x \ d^4 x'  \bigg[
  \left< J^{\mu-}({x})  J^{\nu+}({x}')
\right>_o
 \tilde{D}^{ret}_{\mu \nu}({x}, {x}')
  +\frac{i}{4}  \left< J^{\mu-}( {x})  J^{\nu-}({x}')
\right>_o \tilde{D}^{H}_{\mu \nu}({x}, {x}')  \bigg]
\end{eqnarray}
while the term linear in $\xi^j$ vanishes. The quadratic term in
$S^E_{IF}$ reduces to

\begin{eqnarray}
\label{ }
\left< \left< ( S^E_{IF} )^2 \right>_o \right>_\xi=
  \int d^4 x \ d^4 x' \ d^4 y \ d^4 y'
  \left< J^{\mu-}({x})  J^{\alpha-}({y})
\right>_o
 \tilde{D}^{ret}_{\mu \nu}( {x}, {x}')
\tilde{D}^{ret}_{\alpha \beta}( {y}, {y}')
\kappa^\nu_i  \kappa^\beta_j
\left<
\xi^i({x}')
 \xi^j( {y}')
 \right>_\xi
\end{eqnarray}
where we've dropped higher order terms that enter at $\mathcal{O}(q^3 )$. Thus, at leading order the influence force takes the form

\begin{equation}
\label{force}
f_k(\tau)= \frac{\delta}{\delta z^{k-}(\tau)} \left[\left< \left< S^E_{IF} \right>_o \right>_\xi +\frac{i}{2} \left< \left< ( S^E_{IF} )^2 \right>_o \right>_\xi \right].
\end{equation}
The explicit form for (\ref{force}) can be obtained after an integration by parts and evaluating the expectation values of the atom's current density
separating into three distinct contributions $f_k = f_{1k} + f_{2k} + f_{3k}$. 

The first component arises from the intrinsic fluctuations of the field

\begin{eqnarray}
\label{IF1}
f_{1k}(\tau)=\frac{q^2}{2} \int_{t_{i}}^{t_f} d\lambda \ \delta^{ij} & \underbrace{ g_{ret}(\tau, \lambda) } \partial_k(x)  \underbrace{  \mathcal{G}^{H}_{ij}(x , z^\alpha(\lambda)) } &   \bigg|_{x = z(\tau)}.
\\
& \sim \alpha \ \ \ \ \ \ \ \ \ \ \ \ \sim \left< \Delta \vec{E}^2 \right> & \nonumber 
\end{eqnarray}
The retarded Green's function for the atom $g_{ret}$ quantifies its response to an external field and so represents the atom's dynamic polarizability, $\alpha$. The Hadamard function, $ \mathcal{G}^{H}_{ij}$, describes electric field fluctuations and is constructed by contracting each index of $\tilde{D}^{\mu \nu}_H$ with $\kappa^i_\mu$.   Therefore $f_{1k}$ 
can be related to the recognizable form for the interaction energy of a polarizable body with the electromagnetic field $U_{int} = \frac{1}{2} \alpha \vec{E}^2$. 

The second component follows from dipole moment fluctuations and is the analog of the London force for the atom-surface geometry. Heuristically, we can construct this component of the force by considering the interaction energy of a quantum dipole in front of a dielectric surface. First, the atom is brought near the surface leading to the induction of an image dipole and therefore an image field. The interaction energy of this system is then given by $U_{int}  = - \vec{p} \cdot \vec{E}_{image}$ where $\vec{p}$ is the atom's instantaneous dipole moment and $\vec{E}_{image}$ is the electric field from its image. The image field can be expressed as the convolution of the classical electromagnetic Green's function for the electric field, $\mathcal{G}^{jk}_{ret} $, with the image dipole $E^j_{image} = \int d^4 x' \ \mathcal{G}^{jk}_{ret} \ p^{image}_k$ finally taking the expectation value of the interaction energy we find $U_{int} = - \int d^4 x \ \left<   p_j \   p^{image}_k  \right> \mathcal{G}^{jk}_{ret} $. Because the atom and its image's internal degrees of freedom are correlated  we find that the expectation value of the dipole moments of the atom and image relate directly to the fluctuations of the atom's dipole moment alone $2 \left<   p_j \   p^{image}_k  \right>  \sim  g_H  = \left<   \{ Q_j ,   Q_k  \} \right>  $ showing the relation between our heuristic derivation and the exact expression below

\begin{eqnarray}
\label{IF2}
f_{2k}(\tau)=\frac{q^2}{2} \int_{t_{i}}^{t_f} d\lambda \ \delta^{ij} g_{H}(\tau, \lambda) \partial_k(x) \mathcal{G}^{ret}_{ij}(x, z^\alpha(\lambda)) \bigg|_{x = z(\tau)}.
\end{eqnarray}
The last component of the force, $f_{3k}$, is similar to the first with the exception that the force arises from the \textit{induced} fluctuations of the field. We note from the semi-classical equation of motion for the field 
that the electric field induced by fluctuations within the medium takes the form $E^k_{ind} = \int_V d^4 x' \ \mathcal{G}^{kj}_{ret} \ \xi_j$. By simply taking the expression for $f_{1k}$ and substituting the Hadamard function by the two-point function for \textit{induced} fluctuations we find

\begin{eqnarray}
\label{}
f_{3k}(\tau) \rightarrow \frac{q^2}{2} \int_{t_{i}}^{t_f} d\lambda \ \delta_{kj}  { g_{ret}(\tau, \lambda) } \partial_k(x) \left< \{  E^k_{ind}(x), E^j_{ind}(z^\alpha(\lambda)) \}  \right>    \bigg|_{x = z(\tau)}.
\nonumber 
\end{eqnarray}
which when expanded out takes the form

\begin{eqnarray}
\label{IF3}
f_{3k}(\tau)=    \frac{q^2}{2} \int_{t_{i}}^{t_f} d \lambda \int_V d^3 x \int_{t_{i}}^{t_f} dt \int_{t_{i}}^{t_f} dt' \ g_{ret}(\tau, \lambda) 
\tilde{G}^{med}_H(t,t') \mathcal{G}^{ij}_{ret}(z^\alpha(\lambda), x) \partial_k(x') \mathcal{G}^{ret}_{ij}({x}'; t', \vec{x})\bigg|_{{x}'={z}(\tau)}
\end{eqnarray}
where the collective effect of the medium's fluctuations is accounted for in

\begin{equation}
\label{ }
\tilde{G}^{med}_H(t,t') = \tilde{g}_H(t,t')  +  \int_{t_{i}}^{t_f} dt_1 \int_{t_{i}}^{t_f} dt_2 \  \tilde{g}_{ret}(t,t_1) 
  \tilde{g}_{ret}(t', t_2) \frak{G}_H(t_1, t_2).
\end{equation} 
So we find that at leading order in perturbation theory the origin of each force component is due to the quantum and thermal fluctuations in a different degree of freedom. 

Equations (\ref{IF1}), (\ref{IF2}) and (\ref{IF3}) which form the centerpiece of this paper, provide the fully dynamical and self-consistent force between an atom and a general medium including arbitrary geometry and composition \cite{FN2}
%\footnote{arbitrary composition means that we are free to choose the spectral density of the reservoir} 
for the dielectric medium.

%%%%%%%%%%%%%%%%%%%%%%%%%%%%%%%%%%
\section{Nonequilibrium Atom-Surface Force}
%%%%%%%%%%%%%%%%%%%%%%%%%%%%%%%%%%

In this section we will study the force between an atom situated in vacuum ($z > 0$) and a  dielectric half-space occupying the region ($z < 0$). Our goal here is to find the kernels of the electromagnetic field. The retarded Green's function for the field in a lossy dielectric half-space is well known, and so our focus in this section will be to calculate the Hadamard function quantifying the intrinsic fluctuations of the field. To do this we must find the solution for the time dependent field operator by solving the equation (\ref{HEOM}). 
We begin by taking the divergence of the Heisenberg equation of motion for the homogeneous solution, $\hat{\vec{A}}_o(s, \vec{x})$, for the field (\ref{HEOM}) giving 

\begin{equation}
\label{CE}
\nabla \cdot  \int_{t_i}^{t_f} dt' \ \varepsilon(\dot{t},t', \vec{x} )  \hat{\vec{A}}_o(\dot{t'}, \vec{x}) = 0
\end{equation}  
where $\varepsilon(t,t')$ is the dynamical permittivity and an over dot denotes a time derivative.  

Because the temporal gauge is incomplete the vector potential can be augmented by the gradient of any time-independent scalar without altering the electric or magnetic fields, the condition provided by
(\ref{CE}) fixes the residual gauge freedom giving equivalent physics to the generalized Coulomb gauge \cite{Eberlein06}.

\begin{equation}
\label{ }
\hat{A}^0 = 0 \ \ \ \ \ \  \nabla \cdot \int_{t_i}^{t_f} dt' \ \varepsilon(\dot{t},t', \vec{x}) \hat{\vec{A}}(\dot{t'}, \vec{x}) = 0
\end{equation}
For a piecewise uniform permittivity the field becomes transverse within either half-space. 

For the study of nonequilibrium dynamics of our system we are restricted to compact intervals for the time integrations appearing in the actions describing the total system. 
%Thus, we need to adopt the Laplace transform to solve (\ref{HEOM}) which accounts for the initial conditions.
Using the Laplace transform we can simplify the integrodifferential equation (\ref{HEOM}) for compact time intervals which results in the transformed equation

\begin{equation}
\label{}
  - \nabla^2  \hat{A}_o^j(s, \vec{x})  +   \varepsilon( i s, \vec{x}) s^2  \hat{A}_o^j(s, \vec{x})   = \phi^j_o(s, \vec{x})  = \dot{\hat{A}}^j(t=t_i, \vec{x})+   \varepsilon( i s, \vec{x}) s \hat{A}^j(t=t_i, \vec{x})
 \end{equation}
where $\varepsilon( \omega, \vec{x} )$ is the frequency dependent permittivity $\varepsilon( i s , \vec{x})= 1 + \Omega_P^2 \tilde{g}_{ret}( i s, \vec{x})$ and $\phi^i_o(s, \vec{x})$ encapsulates the effect of the initial conditions. 
We assume that the field and the medium only begin to interact at the initial time $t_i$ and that just prior to $t = t_i$  the field can be found in a plane wave expansion 

\begin{equation}
\label{ffe}
\hat{A}^j(t_i^-,\vec{x}) = \sum_\sigma \int d^3 k \sqrt{ \frac{1}{2 (2 \pi)^3 k }} \bigg[ e^j_\sigma(\vec{k})  a_\sigma (\vec{k})  e^{i \vec{k} \cdot \vec{x} } + h.c. \bigg]
\end{equation}
where $k = | \vec{k}|$, $e^j_\sigma(\vec{k})$ is the $j$th component of a $\sigma$-polarized wave in the Coulomb gauge and $a_\sigma (\vec{k})$ is the destruction operator for a photon of wave vector $\vec{k}$ and polarization $\sigma$. In the Coulomb gauge $\sigma$ is summed over $TE$ (transverse electric) and $TM$ (transverse magnetic) polarization.
By taking the appropriate derivatives and reading off the Fourier amplitudes of (\ref{ffe}) we can identify $\phi^i_o(s, \vec{k})$ as

\begin{equation}
\label{ }
\hat{\phi}^j_o(s, \vec{k}) = \sum_\sigma  \sqrt{ \frac{(2 \pi)^3}{2  k }} (\varepsilon( i s, \vec{x}) s - i k) e^j_\sigma(\vec{k}) a_\sigma (\vec{k}).
\end{equation}

We now consider plane wave solutions originating from either half-space. For points within the dielectric we find 

\begin{equation}
\label{A_med}
\hat{A}^{j-}(s, \vec{k}) = \frac{\hat{\phi}^j_o(s, \vec{k})}{\varepsilon(is) s^2 + k^2}  
\end{equation}
and for points in vacuum 

\begin{equation}
\label{A_vac}
\hat{A}^{j+}(s, \vec{k}) = \frac{\hat{\phi}^j_o(s, \vec{k})}{ s^2 + k^2}
\end{equation}
where + (-) in this section denotes waves in vacuum (dielectric) and in (\ref{A_vac}) the permittivity is taken to $1$ in $\hat{\phi}^j_o(s, \vec{k})$. 

Inverting the Laplace transform gives the time dependence of the  field amplitude for planes waves with momentum $\vec{k}$. In vacuum we have 

\begin{equation}
\label{ }
\hat{A}^{j+}(t, \vec{k}) =  \sum_\sigma  \sqrt{ \frac{(2 \pi)^3}{2  k }}  e^j_\sigma(\vec{k}) a_\sigma (\vec{k}) e^{-i k (t-t_i)} \theta(t-t_i)
\end{equation}
and within the dielectric medium we have the general result 

\begin{equation}
\label{A_med}
\hat{A}^{j-}(t, \vec{k}) = \frac{1}{2 \pi i} \int_{\mathcal{C}} ds \ e^{s(t-t_i)} \frac{\hat{\phi}^j_o(s, \vec{k})}{\varepsilon(is) s^2 + k^2}  
\end{equation}
which depends upon the specific choice of matter-reservoir coupling. For Ohmic reservoir spectral density the dispersion relation has four distinct poles and the Bromwich integral can 
be inverted exactly

\begin{equation}
\label{A_med_Ohmic}
\hat{A}^{j-}(t, \vec{k}) = \sum_{l} \sum_\sigma  \sqrt{ \frac{(2 \pi)^3}{2  k }}  e^j_\sigma(\vec{k}) a_\sigma (\vec{k}) r_l e^{ s_l (t-t_i)} \theta(t-t_i)
\end{equation}
where $r_l$ is the residue of $(\varepsilon(is)s-ik)/(\varepsilon(is) s^2 + k^2)$ from the $l$th pole and $s_l$ is the $l$th pole. 

Given the plane wave solutions above we can now construct the total field by considering right and left-incident waves. 
Waves that approach the vacuum-dielectric interface from the right reflect and transmit. The sum of the incident, reflected and transmitted components gives the total solution where the boundary conditions on the field at the interface 
\begin{eqnarray}
\label{ }
E_{\|}(0^+) = E_{\|}(0^-) \ \ \ \ \ \text{and} \ \ \ \ \  D_{z}(0^+) = D_{z}(0^-)
\\
H_{\|}(0^+) = E_{\|}(0^-) \ \ \ \ \ \text{and} \ \ \ \ \  B_{z}(0^+) = B_{z}(0^-)
\end{eqnarray}
determine the Fresnel transmission and reflection coefficients. For example, in the case of TE waves the electric field is parallel to the vacuum-dielectric interface and because the 
electric field is parallel to the vector potential the $x$ and $y$ components of the vector potential are continuous across $z = 0$ leading to the relation (see \ref{AR}) $1 + R^R_{TE} = T^R_{TE}$ (to avoid confusion we stress here that this is a constraint on the field amplitude and not on the energy). Because our dielectric model does not have a magnetic response all components of the magnetic field are continuous across $z=0$. For the case of TE waves the continuity of the $x$ and $y$ components of the magnetic field give $k_z(1-R^R_{TE}) = K_z T^R_{TE}$ where $K_z$ is chosen so that the transmitted component of the field satisfies the wave equation in the medium. Following the same procedure for TM polarized waves defines the remaining reflection and transmission coefficients.
This allows us to express the right-incident waves as

\begin{eqnarray}
\label{AR}
\hat{A}^{j}_{R}(x) =    \sum_{\sigma=TE,TM}  \int d^3 k \  \frac{1}{\sqrt{2(2\pi)^3 k }}  \theta(-k_z) e^{ -i k(t -t_i) + i \vec{k}_{\|} \cdot \vec{x}_{\|}} 
 a^R_{ \sigma} (\vec{k}) e^j_{\sigma} \theta(t-t_i)
  \nonumber \\
\times
\bigg[ 
 \bigg(
e^{i k_z z }   + R^R_\sigma e^{- i k_z z}  \bigg)  \theta(z)
+
T^R_\sigma e^{i K_z z }   \theta(-z)  \bigg] + h.c..
\end{eqnarray}
where the Fresnel coefficients become

\begin{eqnarray}
\label{ }
& R^R_{TE}  =  \frac{k_z - K_z}{k_z + K_z} \ \ \ \ \  T^R_{TE}  =  \frac{2 k_z }{k_z + K_z}  \ \ \ \ \ R^R_{TM}  =   \frac{ \varepsilon(k)k_z -  K_z }{ \varepsilon(k) k_z + K_z}  \ \ \ \ \  T^R_{TM}  =   \frac{ 2 \sqrt{\varepsilon( k )} k_z }{ \varepsilon(k) k_z + K_z}. &
\nonumber 
\end{eqnarray}
Here $a^R_\sigma(\vec{k})$ is the destruction operator for left-moving waves with wave vector $\vec{k}$ ($k_z < 0$) and polarization $\sigma$, the z-component of the wave vector in the medium is $K_z =   \text{sign}(k_z) \sqrt{ \varepsilon( k) k^2 - k^2_\|}$, and the polarization vectors can be conveniently expressed in differential form as $\hat{e}_{TE} = (- \Delta_\|)^{-1/2} ( -i \partial_y, i \partial_x, 0)$ and  $\hat{e}_{TM} = (\Delta \Delta_\|)^{-1/2} ( - \partial_x \partial_z,  - \partial_y \partial_x, \Delta_\|)$.

Likewise, we can follow the same procedure for left-incident waves. For a general medium one would employ $(\ref{A_med})$, in the case of Ohmic reservoir spectral density we find  

\begin{eqnarray}
\label{AL}
\hat{A}^{j}_{L}(x) =     \sum_{\sigma = TE, TM} \sum_l \int d^2 k \int d K_z \  \frac{ 1}{\sqrt{2(2\pi)^3 k' }}  \theta(K_z) r_l e^{s_l(t -t_i) + i \vec{k}_{\|} \cdot \vec{x}_{\|}} 
  a^L_{ \sigma} (\vec{k}') e^j_{\sigma} \theta(t-t_i)
  \nonumber \\
\times
\bigg[ 
 \bigg(
e^{i K_z z }   + R^L_\sigma e^{- i K_z z}  \bigg)  \theta(-z)
+
T^L_\sigma e^{i k_{l,z} z }   \theta(z)  \bigg] + h.c.
\end{eqnarray}
where the Fresnel coefficients are

\begin{eqnarray}
\label{ }
& R^L_{TE}  =   \frac{K_z - k_{l,z} }{K_z + k_{l,z} } \ \ \ \ \  T^L_{TE}  =  \frac{2 K_z }{k_{l,z} + K_z} \ \ \ \ \  R^L_{TM}  =   \frac{ K_z -  \varepsilon( is_l )k_{l,z}  }{  K_z+ \varepsilon( is_l ) k_{l,z} }  \ \ \ \ \  T^L_{TM}  =   \frac{ 2 \sqrt{\varepsilon(  is_l )} K_z }{ \varepsilon( is_l ) k_{l,z} + K_z}  &
\nonumber 
\end{eqnarray}
the wave vector in the medium is $\vec{k}' = (\vec{k}_\|, K_{z})$ with magnitude $k' = |\vec{k}' |$,
and $k_{l,z}$ is chosen to solve the wave equation in vacuum $k_{l,z} =  \sqrt{ - {s_l}^2 - k^2_\|}$ with $\text{Im} [ k_{l,z} ] > 0 $.
By considering the cases where the dielectric becomes an ideal conductor ($\varepsilon \rightarrow \infty$), $\varepsilon$ becomes a real constant, or the dielectric becomes transparent ($\varepsilon = 1$) one can verify that the expressions for the field above reduce to the expected half-space \cite{Eberlein06} and vacuum forms (after choosing an appropriate initial state).

We can now construct the two-point functions of the field under the influence of the medium. The retarded Green's function solves (\ref{GFeqn})
%whose solution can be taken from (\ref{AL}) and (\ref{AR}) where $ \hat{\phi}^l_o(s, \vec{k})$ is replaced with $\delta^{lm} e^{ -s (t' - t_i)- i \vec{k} \cdot \vec{x}'}$.
and can be split into two pieces; one corresponding with the source point in vacuum and the other with source point in the medium. This can be easily understood by noting that $\hat{A}^{R,L}$ is  proportional to $a^{R,L}_\sigma$ and therefore $\left< A^R A^L \right> = 0$.  Therefore we can decompose $\tilde{D}^{ij}_{ret}$ into a portion due to right and left-incident waves

\begin{equation}
\label{ }
\tilde{D}^{ij}_{ret}(x, x') = \tilde{D}^{ij}_{ret,R}(x, x') + \tilde{D}^{ij}_{ret,L}(x, x'). 
\end{equation}
The retarded Green's function gives the electric field from a point dipole located at $\vec{r}$

\begin{equation}
\label{ }
E^{dip}_i(x) =   \int d t' \mathcal{G}_{ij}^{ret}( x; t' ,\vec{r}) p^j_o(t').
\end{equation}
Dipole oscillations will lead to emitted radiation that will be seen at $x$ at retarded times. 
By noting that the source point for the retarded Green's function appearing in the component of the atom-surface force due to medium fluctuations lies in $V$ and the observation point lies to the future in the vacuum region we can ascertain that information about medium fluctuations is contained entirely in the left-incident component of $\mathcal{G}^{ij}_{ret}$ which allows the replacement $ \mathcal{G}^{ret}_{ij} \rightarrow  \mathcal{G}^{ret,L}_{ij}$ in (\ref{IF3}).

The Hadamard function for the field can be constructed from
the solution for the field operators (\ref{AR}) and (\ref{AL}) 
which also decomposes into a left and right-incident component

\begin{equation}
\label{ }
\tilde{D}^{ij}_H(x, x') = \left< \{ \hat{A}^{i}_L(x) ,\hat{A}^{j}_L(x') \} \right> +  \left< \{ \hat{A}^{i}_R(x) ,\hat{A}^{j}_R(x') \} \right> .
\end{equation}
The crucial point we would like to make now is that for a dissipative medium the intrinsic fluctuations of the field are exponentially suppressed in time for fluctuations emanating from within the medium. 

The time dependence of the field for wave vector $\vec{k}$ emanating from within the medium can be calculated by inverting the Laplace transform in (\ref{A_med}).
Because all of the poles (and/or branch cuts) in the dispersion relation lie to the left of the imaginary axis in the complex $s$-plane the intrinsic fluctuations of the field associated with left-incident waves are damped away in time so that at late times 

\begin{equation}
\label{ }
\tilde{D}_{ij}^{H}(x, x') \stackrel{t\rightarrow \infty}{ \rightarrow}  \left< \{ \hat{A}_{i}^R(x) ,\hat{A}_{j}^R(x') \} \right>  \equiv  \tilde{D}^{H,R}_{ij}(x, x').
\end{equation}

%%%%%%%%%%%%%%%%%%%%%%%%%%%%%%%%%%
\subsection{Long-Time Limit} 
%%%%%%%%%%%%%%%%%%%%%%%%%%%%%%%%%%

We now specialize to the case of a stationary atom and take the long-time limit. Because this system exhibits dissipation we expect that a steady-state exists at late times. In the previous section we have already noted that for general reservoir-matter coupling that field fluctuations emanating from inside the dielectric medium will be exponentially suppressed in time. Within the medium the matter fluctuations will be defined by the reservoir at late times as the polarization field is thermalized. The collective fluctuations of the medium is described by the kernel $\tilde{G}^{med}_H(t,t')$ containing components due to the intrinsic fluctuations of the matter and reservoir. We showed explicitly for Ohmic reservoir spectral density that the intrinsic fluctuations of the matter damp away in time (\ref{fluc}) and will for general reservoir-matter coupling. Therefore as $ t_i  \rightarrow -\infty$ we find 

\begin{equation}
\label{ }
\tilde{G}^{med}_H(t,t')  \rightarrow  \int_{ -\infty }^{ \infty } dt_1 \int_{ -\infty }^{ \infty } dt_2 \  \tilde{g}_{ret}(t,t_1) 
  \tilde{g}_{ret}(t', t_2) \frak{G}_H(t_1, t_2).
\end{equation} 
where the upper limit of the time integrals can be taken to infinity because the matter's retarded Green's functions contain theta functions which restrict the integration range over past times. By taking the Fourier transform we find  an equivalent description in frequency space

\begin{equation}
\label{ }
\tilde{G}^{med}_H( \omega' )  \rightarrow   \tilde{g}^*_{ret}(\omega') 
  \frak{G}_H( \omega' ) \tilde{g}_{ret}( \omega' )
\end{equation} 
The retarded Green's function for the matter satisfies (\ref{MEOM}) sourced by a delta function

\begin{equation}
\label{ }
\frac{\partial^2}{\partial t^2}  \tilde{g}_{ret}(t, t') + \omega^2 \tilde{g}_{ret}(t, t') - \int_{t_i}^{t_f} dt'' \frak{G}^{ret}(t,t'')\tilde{g}_{ret}(t'', t') = \delta(t-t')
\end{equation}
whose Fourier transform gives (again $t_f , - t_i \rightarrow \infty$)

\begin{equation}
\label{ft_MEOM}
-{\omega'}^{2}  \tilde{g}_{ret}( \omega' ) + \omega^2 \tilde{g}_{ret}(  \omega' ) - \frak{G}^{ret}(  \omega' )\tilde{g}_{ret}( \omega' ) = 1.
\end{equation}
where $\tilde{g}_{ret}( \omega' )$ is a complex function of frequency. Multiplication of (\ref{ft_MEOM})  by $\tilde{g}^*_{ret}( \omega' )$ and then taking the imaginary part provides an identity between $\frak{G}^{H}$ and $\tilde{g}^{ret}$ \cite{Eckhardt82,Eckhardt84}

\begin{equation}
\label{ }
 \tilde{g}^*_{ret}( \omega' ) \text{Im} [ \frak{G}^{ret}(  \omega' )  ] \tilde{g}_{ret}( \omega' ) = \text{Im} [ \tilde{g}_{ret}( \omega' ) ].
\end{equation}
Using the fluctuation-dissipation relation (\ref{FDR_RM}) we find 

\begin{equation}
\label{ }
 \tilde{g}^*_{ret}( \omega' )\frak{G}^{H}(  \omega' )  \tilde{g}_{ret}( \omega' ) = 2 \coth(\beta_X \omega'/2) \text{Im}  [\tilde{g}_{ret}  ( \omega' ) ]= \tilde{G}^{med}_H(  \omega'  )
\end{equation}
where the last equality holds only for $t_i \rightarrow - \infty$.
The imaginary part of $\tilde{g}_{ret} (\omega')$ relates directly to the imaginary part of the permittivity so that in the long-time limit
we find the asymptotic behavior
 
 \begin{eqnarray}
\label{IF_LT}
f_k = \frac{q^2}{4 \pi} \int_{-\infty}^{\infty} d \omega' \ \delta^{ij} [ g^*_{ret}( \omega') \partial_k(x) \mathcal{G}^{H, R}_{ij}( \omega' , \vec{x}, \vec{z})+g^*_{H}( \omega') \partial_k(x) \mathcal{G}^{ret}_{ij}( \omega', \vec{x}, \vec{z} )]_{x = z(\tau)}
\nonumber \\
+\frac{q^2}{2 \pi } \int_{-\infty}^{\infty} d \omega'   \int_V d^3 x  \
 \text{Im}[ \varepsilon(\omega') ]
\coth (\beta_X \omega' /2) g^*_{ret}(  \omega' ) 
 \mathcal{G}^{ij*}_{ret, L}(  \omega', \vec{z}, \vec{x}) \partial_k(x') \mathcal{G}^{ret,L}_{ij}(  \omega', \vec{x}'; \vec{x})\bigg|_{{x}'={z}(\tau)}.
\end{eqnarray}

\subsubsection{Fluctuation-Dissipation Relation for the Field in Vacuum}

In the long time limit, when a steady-state has been attained, we can study the features of right-incident waves, $A^R$, in terms of Fourier modes in the vacuum region as in 
(\ref{AR})

\begin{eqnarray}
\label{}
\hat{A}^{j}_{R}(x) =    \frac{1}{\sqrt{2(2\pi)^3  }}    \sum_{\sigma=TE,TM}  \int \frac{ d^3 k }{k}  \   \theta(-k_z)  
  \bigg[
 a^R_{ \sigma} (\vec{k}) e^j_{\sigma} e^{- i \omega (t -t_i) + i \vec{k} \cdot \vec{x} } \bigg(
1  + R^R_\sigma e^{- i 2 k_z z}  \bigg)  + h.c. 
  \bigg] \theta(z) .
\end{eqnarray}
The Wightman function for the field provides all information about propagation and fluctuations of right-incident waves in the vacuum region (evaluated here at coincident spatial arguments for later convenience)

\begin{eqnarray}
\label{}
\tilde{D}_R^{jl +} (x; t', \vec{x}) = \left<\hat{A}^{j}_{R}(x)\hat{A}^{l}_{R}(t', \vec{x} ) \right> =      \frac{1}{2(2\pi)^3 }    \sum_{\sigma=TE,TM} \int \frac{ d^3 k }{k} \    \theta(-k_z)  
\bigg[ e^j_{\sigma} e^{ i \vec{k} \cdot \vec{x} } \bigg(
1  + R^R_\sigma e^{- i 2 k_z z}  \bigg)
\bigg] 
  \nonumber \\
    \times
\bigg[  e^{l*}_{\sigma} e^{-i \vec{k} \cdot \vec{x} } \bigg(
1  + R^{R*}_\sigma e^{ i 2 k_z z}  \bigg) \bigg] 
  \bigg[ \coth \beta_E k/2 \cos k(t-t') - i \sin k(t-t')
  \bigg]
  \theta(z)
\end{eqnarray}
where we have assumed that the right-incident component of the field is in a thermal state at inverse temperature $\beta_E$. 
Two times the real part gives the Hadamard function and two times the imaginary part relates to the retarded propagator

\begin{equation}
\label{ }
\tilde{D}^{jl}_{H,R} (x; t', \vec{x})  = 2 \text{Re}[ \tilde{D}^{jl +}_R (x; t', \vec{x})]  \ \ \ \ \ \ \ \ \ \  \tilde{D}^{jl}_{ret,R} (x; t', \vec{x})  = 2 i \theta(t-t')  \text{Im} [ \tilde{D}^{jl +}_R (x; t', \vec{x}) ].
\end{equation}
After taking the Fourier transform of these stationary kernels with respect to $t-t'$ we find that they are related by the fluctuation-dissipation relation

\begin{equation}
\label{ }
\tilde{D}^{jl}_{H,R} (\omega', \vec{x}, \vec{x})  =  2 \coth (\beta_E \omega' /2) \ \text{Im} [ \tilde{D}^{jl}_{ret,R} (\omega', \vec{x}, \vec{x}) ]
\end{equation}
at the same point.

%%%%%%%%%%%%%%%%%%%%%%%%%%%%%%%%%%
\subsection{Steady-State Atom-Surface Force}
%%%%%%%%%%%%%%%%%%%%%%%%%%%%%%%%%%

For comparison with previous works we note that $\mathcal{G}^{ij}_{ret}({\omega'},
\vec{x}, \vec{x}')$ and $ q^2 g_{ret}({\omega'})$ in our treatment can
be identified with the dyadic Green's function for the electric
field, $G^{ij}/(4\pi)$, and $4\pi \alpha({\omega'})$ the
frequency-dependent polarizability used in 
\cite{APS05, Ant_JPhys, APSS08}.

By using the fluctuation-dissipation
relation the atom-surface force can be brought into the same form as derived by
others. 
After taking the long time limit,
Fourier transforming in time, assuming global thermodynamic equilibrium and
specifying the atom's trajectory to be static (\ref{IF_LT}) can be used to express 
the force in equilibrium.
This expression can be simplified by the use of an identity. 

If we specify that the spacetime point $x$ lies in vacuum and to the causal future 
of $x'$ with spatial component in $V$ then $\mathcal{G}^{ret,L}_{ik}(x,x')$ satisfies (\ref{GFeqn}) and can be shown to satisfy the identity 

\begin{equation}
\label{GF_id}
\int_{V} d^3 x' \ \delta^{lk} \text{Im} [\varepsilon({\omega'}, \vec{x}') ]  \mathcal{G}^{ij*}_{ret,L}(  {\omega'}, \vec{x}, \vec{x}')  \mathcal{G}^{ret,L}_{kj}(  {\omega'}, \vec{x}, \vec{x}') =  \delta_{kj} \text{Im} [ \mathcal{G}^{ij}_{ret,L}(  {\omega'}, \vec{x}, \vec{x}) ]
\end{equation}
in Fourier space.

%can be used to simplify our expression for the force in equilibrium where the integral is over all space. By choosing a volume integration over the left half-space where the surface of the volume passes just outside of the dielectric (so that surface terms from an integration by parts vanish) and the points $\vec{z}$ and $\vec{x}$ lie in the right half-space we find
%\begin{equation}
%\label{ }
%\int_V d^3 x \ \text{Im} [\varepsilon({\omega'}, \vec{x}) ]  \mathcal{G}^{ij*}_{ret}(  {\omega'}, \vec{z}, \vec{x})  \mathcal{G}^{ret}_{ik}(  {\omega'}, \vec{x}, \vec{x}') =  \text{Im} [  \mathcal{G}^{ij,R}_{ret}(  {\omega'}, \vec{z}, \vec{x}') ]
%\end{equation}
%where $\mathcal{G}^{ij,R}_{ret}(  {\omega'}, \vec{z}, \vec{x}')$ gives information about fluctuations that have transmitted from the dielectric  into vacuum.

For $z > 0$ the fluctuation-dissipation relation requires
$\mathcal{G}^{H,R}_{ij}({\omega'},\vec{z},\vec{z})=2 \coth(\beta {\omega'}/2) \text{Im}
[ \mathcal{G}^{ret,R}_{ij}({\omega'},\vec{z},\vec{z}) ]$  and \\  $g_H({\omega'})= 2 \coth(\beta
{\omega'}/2) \text{Im} [ g_{ret}({\omega'}) ]$.  With these relations (and borrowing the notation of others) we can rewrite
$f_{k}$ as

\begin{eqnarray}
\label{f1}
f_{k}=\frac{1}{\pi} \int_0^\infty d{\omega'} \ \delta^{ij} \coth(\beta {\omega'}/2) \text{Im}[ \alpha({\omega'}) \partial_k(x) G_{ij}({\omega'}, \vec{x},\vec{z})]\bigg|_{\vec{x}=\vec{z}(\tau)}
\end{eqnarray}
reducing it to the form for the atom-surface force as derived from MQED \cite{Ant_JPhys}.

It is interesting to consider now the specific case where the temperature of the medium differs from the field. Using the fact that 

\begin{equation}
\label{ }
 \text{Im}[  \mathcal{G}^{ij,R}_{ret}(  {\omega'}, \vec{z}, \vec{x}') ] +  \text{Im} [ \mathcal{G}^{ij,L}_{ret}(  {\omega'}, \vec{z}, \vec{x}')]  =  \text{Im} [ \mathcal{G}^{ij}_{ret}(  {\omega'}, \vec{z}, \vec{x}') ]
\end{equation}
we can write the force in the form

\begin{eqnarray}
\label{}
f_k = \frac{q^2}{4 \pi} \int_{-\infty}^{\infty} d {\omega'} \ \delta^{ij} \bigg[ 2 \coth(\beta_E {\omega'}/2)  g^*_{ret}( {\omega'}) \partial_k(x) 
 \text{Im} [ \mathcal{G}^{ij}_{ret}(  {\omega'}, \vec{x}, \vec{z})  ]
+g^*_{H}( {\omega'}) \partial_k(x) \mathcal{G}^{ret}_{ij}( {\omega'},  \vec{x}, \vec{z} ) \bigg]_{x = z(\tau)}
\\
+   { \frac{q^2}{2 \pi } \int_{-\infty}^{\infty} d {\omega'}   \int_V d^3 x  \  \text{Im} [ \varepsilon({\omega'}) ] ( \coth (\beta_M {\omega'} /2) 
- \coth (\beta_E {\omega'} /2))
g^*_{ret}(  {\omega'} )
 \mathcal{G}^{ij*}_{ret,L}(  {\omega'}, \vec{z}, \vec{x}) \partial_k(x') \mathcal{G}^{ret,L}_{ij}(  {\omega'}, \vec{x}', \vec{x})\bigg|_{{x}'={z}(\tau)} }  . 
 \nonumber
\end{eqnarray}

The first term is equivalent to the force associated with the atomic energy level shifts derived in \cite{Wylie84},  when the atom is in a thermal state at temperature $T_E$ it gives back the atom-surface force in thermal equilibrium. The last term gives the correction to the force when the medium and field are out of thermodynamic equilibrium. We denote this correction to the force $f^{neq}_k$

\begin{eqnarray}
\label{fneq}
f^{neq}_k=\frac{q^2}{2\pi} \int_{-\infty}^{\infty} d{\omega'} \int_{V} d^3 y \
\delta^{ij}   g^*_{ret}({\omega'}) \partial_k(x)
 \ \text{Im}[\varepsilon({\omega'})]  \big(\coth[\beta_M {\omega'}/2] -\coth[\beta_E {\omega'}/2] \big) \nonumber \\  \times
 \mathcal{G}_{ret,L}^{ij}({\omega'}, \vec{x}, \vec{y} ) \mathcal{G}^{ret,L*}_{ij}({\omega'}, \vec{z}, \vec{y} )
\bigg|_{\vec{x}=\vec{z}} .
\end{eqnarray}

%%%%%%%%%%%%%%%%%%%%%%%%%%%%%%%
\subsection{Nonequilibrium Steady-State Force}
%%%%%%%%%%%%%%%%%%%%%%%%%%%%%%%

In this section we elucidate the physics of the nonequilibrium correction 
to the force by evaluating the convolution of the Green's functions in (\ref{fneq}) 
and comparing the results with the force in equilibrium. After some manipulation 
(\ref{fneq}) can be brought into the form

\begin{eqnarray}
\label{fneq_ew}
f^{neq}_{k}(T_M, T_E)_{EW} = - \frac{  \sqrt{2}}{  \pi} \int_0^\infty d{\omega'} \int_1^\infty dq \ q \ {\omega'}^4 \text{Re}[\alpha({\omega'})] \big(\coth(\beta_M {\omega'}/2)-\coth(\beta_E {\omega'}/2) \big) 
 \sqrt{q^2-1}
 \nonumber \\
 \times 
 \sqrt{\text{Re}[\varepsilon({\omega'})]-q^2 +|\varepsilon({\omega'})-q^2|}
 \bigg[ \frac{1}{|\sqrt{1-q^2}+\sqrt{\varepsilon({\omega'})-q^2} |^2} +\frac{(2q^2-1)(q^2 + |\varepsilon({\omega'})-q^2|)}{|\varepsilon({\omega'}) \sqrt{1-q^2}+\sqrt{\varepsilon({\omega'})-q^2} |^2}  \bigg]  e^{-2z {\omega'} \sqrt{q^2-1}} 
 \end{eqnarray}

\begin{eqnarray}
\label{fneq_pw}
f^{neq}_{k}(T_M, T_E)_{PW} =  \frac{  \sqrt{2}}{  \pi} \int_0^\infty d{\omega'} \int_0^1 dq \ q \ {\omega'}^4 \text{Im}[\alpha({\omega'})] \big(\coth(\beta_M {\omega'}/2)-\coth(\beta_E {\omega'}/2) \big) \sqrt{q^2-1}
\nonumber \\
\times
 \sqrt{\text{Re}[\varepsilon({\omega'})]-q^2 +|\varepsilon({\omega'})-q^2|}
 \bigg[ \frac{1}{|\sqrt{1-q^2}+\sqrt{\varepsilon({\omega'})-q^2} |^2} +\frac{(q^2 + |\varepsilon({\omega'})-q^2|)}{|\varepsilon({\omega'}) \sqrt{1-q^2}+\sqrt{\varepsilon({\omega'})-q^2} |^2}  \bigg]. \ \ \ \ \ \ \ \ \ 
\end{eqnarray}
where the force splits into a component due to evanescent waves ($EW$) and one due to propagating waves ($PW$), see appendix for details
(the properties of the polarizability have been used to simplify the frequency integrations). 
Note in particular that the evanescent component of the nonequilibrium correction is equivalent to the difference between two evanescent
components to the atom-surface force in equilibrium (\ref{PFEWc}). The propagating wave component gives rise to a constant space-independent \textit{wind} term which results from the 
radiation of the field into the vacuum region ($z>0$).  
The imaginary part of the polarizability in (\ref{fneq_pw}) has support in a narrow band around the resonances of the atom, indeed for our oscillator model the imaginary part of the polarizability is proportional to a delta function which allows the frequency integral in (\ref{fneq_pw}) to be done directly.  However, for realistic laboratory temperatures the effect of the propagating waves is exponentially suppressed.
(See  \cite{Ant_JPhys} 
for a detailed derivation of  asymptotic properties of the evanescent wave nonequilibrium contribution to the force.)

We now have an expression for the atom-surface force for a stationary system when the dielectric medium is out of thermal equilibrium with the field. By dissecting the Lifshitz force we can intuitively argue for the form of the force derived previously. First, in thermal equilibrium the Lifshitz force is composed of a propagating and evanescent wave component. The propagating component arises from field fluctuations in the vacuum region and partially transmitted waves from the medium. The evanescent component arises from fluctuations of the field within the dielectric that partially transmit into the vacuum region from incident angles exceeding the critical angle. By subtracting the effect of transmitted waves from the expression for the force in equilibrium, both at temperature $T_E$, we describe the effect of incident and reflected propagating waves in the vacuum region. To account for the waves partially transmitted into the vacuum region we need to add the effect of medium fluctuations within the dielectric at temperature $T_M$ which is done by adding $f^{neq}(T_M, T_E)$ to the equilibrium force. This results in a modification of the evanescent component of the equilibrium force and accounts for the blackbody radiation emitted from the surface when the global system is out of equilibrium  

\begin{equation}
\label{ }
f_k= f_k(T_E) - f^{FF,EW}_k(T_E)+f^{FF,EW}_k(T_M) + f^{neq}_{k}(T_M, T_E)_{PW}.
\end{equation}
%It is interesting to note that the atom-surface force computed with respect to the initially factorized state contains an additional evanescent component arising from fluctuations of the medium.

\begin{figure}[h]
\begin{center}
\includegraphics[width=6.5in]{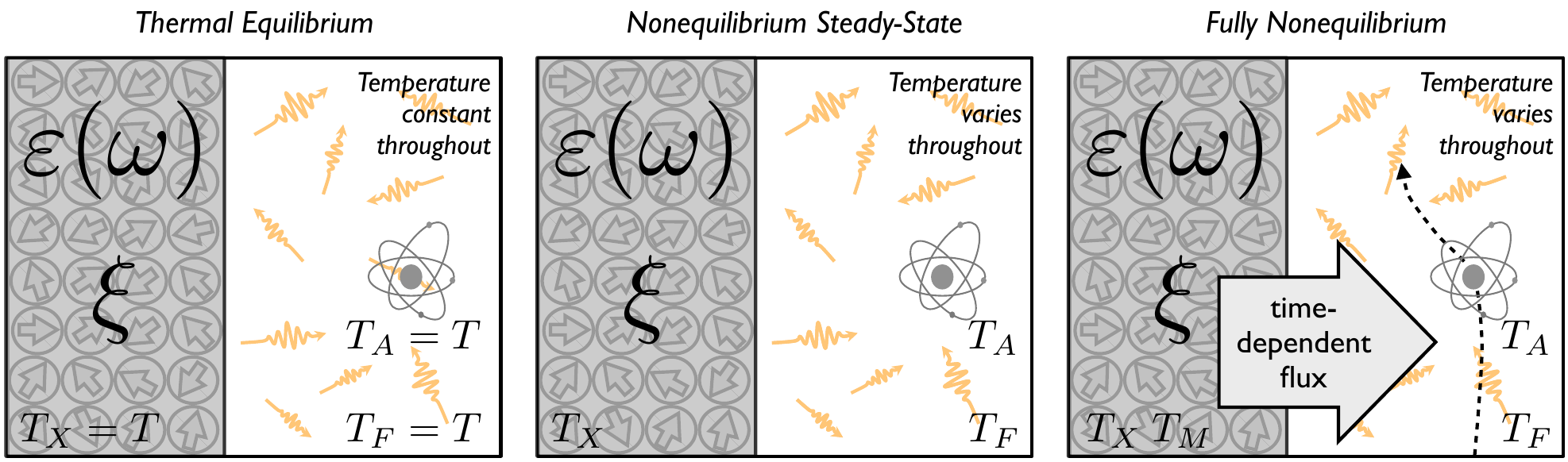}
\label{summ}
\caption{
The figure above summarizes the differences between atom-surface interactions in equilibrium, in a nonequilibrium steady-state, and under fully nonequilibrium conditions.
The grey bulk at left illustrates the matter where the symbols $\varepsilon(\omega)$ and $\xi$ denote that the medium has been coarse-grained and so its averaged effect is described by a permittivity and a stochastic force. The vacuum region at the right of the matter contains an atom and the field. The simplest case is \textit{thermal equilibrium} where are all parties are described by a thermal state at temperature $T$, the long time limit is taken, and the position of the atom is fixed. The equilibrium case can be generalized by the consideration of \textit{nonequilibrium steady-states} where all parties are described by thermal states (e.g. $T_A$, $T_F$, and $T_X$ are the atom's, field's and reservoir's temperature), the atom is held fixed, and the long-time limit is taken so that the state of the entire system no longer evolves in time. A \textit{fully nonequilibrium} treatment describes dynamics. For this case the total system is time evolved from an initial state given above as thermal for each party. The illustration depicts the time dependent flux of energy emanating from the medium as well as atom motion. 
}
\end{center}
\end{figure}

%%%%%%%%%%%%%%%%%%%%%%%%%%%%%%%%%%
\subsection{Force as a function of time} 
%%%%%%%%%%%%%%%%%%%%%%%%%%%%%%%%%%

Now that we have isolated the steady-state form for the force we can
identify the components giving rise to force dynamics. We showed previously 
that as time evolves the reservoir will thermalize all parties that it interacts with
eventually  resulting in a steady-state. The power of our nonequilibrium formulation 
is that we can follow the time-evolution of the force from an initial state. 

To highlight these effects, not obtainable through MQED or Lifshitz theory,  
we isolate the dynamical terms from the total force which at early times  
will lead to different predictions from the standard equilibrium theories. 
These corrections come in two terms. First, we showed earlier that the left-incident fluctuations of the field 
are damped away in time due to interaction with the medium. For times shorter than the inverse dissipation rate 
left-incident field fluctuations augment the equilibrium force by

\begin{equation}
\label{fdyn1}
f^{dyn}_{1k}(\tau) =  \frac{q^2}{2} \int_{t_{i}}^{t_f} d\lambda \ \delta^{ij}  g_{ret}(\tau, \lambda)  \partial_k(x)  \mathcal{G}^{H,L}_{ij}(x , z^\alpha(\lambda))    \bigg|_{x = z(\tau)}.
\end{equation} 
The second contribution comes from the induced fluctuations of the field which at early times, before the matter is thermalized by the reservoir, 
gives the additional force contribution

\begin{equation}
\label{ }
f^{dyn}_{2k}(\tau) =  \frac{q^2}{2} \int_{t_{i}}^{t_f} d \lambda \int_V d^4 x  \int_{t_{i}}^{t_f} dt' \ g_{ret}(\tau, \lambda) 
\tilde{g}^{med}_H(t,t') \mathcal{G}^{ij}_{ret,L}(z^\alpha(\lambda), x) \partial_k(x') \mathcal{G}^{ret,L}_{ij}({x}'; t', \vec{x})\bigg|_{{x}'={z}(\tau)}.
\end{equation}
An illustrated summary of the different cases treated in this paper can be found in Figure[1].

\subsubsection{Time dependent force due to field thermalization by eddy currents}

As an explicit example of force dynamics we'll compute the time-dependent contribution to the atom-surface force arising from the thermalization of the field by the medium. For this purpose we'll work explicitly with the Drude model which has been chosen for its simplicity. In addition we'll assume that the field is initially described by a thermal state, factorized from the remaining parties, where the matter is assumed to have already been thermalized by the reservoir. 

The Drude model dispersion relation contains two dissipative oscillating modes, the bulk plasmons, and a third diffusive mode arising from eddy currents \cite{Jackson62}. For weak dissipation (dissipation rate much smaller than the plasma frequency) the plasmons give rise to a constant space independent force at leading order corresponding to black body radiation into vacuum. In distinction the eddy currents establish an evanescent field which varies with atom-surface spacing and so will be the focus of our investigation. After computing the Hadamard function for the field in (\ref{fdyn1}) we find the explicit expression for the dynamical force below

\begin{eqnarray}
\label{}
f^{dyn}_{1k}(\tau) =  - \frac{1}{8 \pi^2} \alpha_o \Omega \sum_{l,l'} \sum_\sigma \int_{0}^{\tau} d\lambda \ \int_0^\infty d k \int_0^1 dq \ k \coth(\beta k/2)  \sin \Omega (\tau - \lambda)
 \nonumber 
 \\
 \times \text{Im} \bigg[ k_{z,l}  \ s_l \ s^*_{l'} \ r_l \ r^*_{l'} \ e^{ s_l \tau + s^*_{l'} \lambda + i (k_{z,l} - k^*_{z,l'} ) z} 
T^L_{\sigma,l} T^{L*}_{\sigma,l'} e_{\sigma,l} \cdot e^*_{\sigma,l'}
\bigg]  \ \ \ \ \ 
\end{eqnarray} 
where $\alpha_o$ is the oscillator's static polarizability, we've set the initial time to zero and we've made the change of variables $K_z = k q$. In the limit of weak dissipation the root, $s_l$, the $z$-component of the wave-vector, $k_{z,l}$, and the residue, $r_l$, for the diffusive mode can be approximated by 

\begin{equation}
\label{ }
s_{eddy} \approx - \frac{\gamma k^2}{k^2 + \Omega_P^2} \ \ \ \ \ r_{eddy}   \approx  \frac{-  k^2 + \Omega_P^2 }{ 2 ( k^2 + \Omega_P^2)}  \ \ \ \ \   k_{z,eddy}   \approx   i k \sqrt{1-q^2}.
\end{equation}
After making the change of variables $ y = \sqrt{1 - q^2}$ we note that the exponential factor $\exp \{   s_l \tau + s^*_{l'} \lambda + i (k_{z,l} - k^*_{z,l'} ) z \} $ suppresses large values of $y$ and $k$ for $ \Omega_P z >>1$ at finite times.  After expanding the integrand for small $y$ and $k$ we find

\begin{eqnarray}
\label{ }
f^{dyn}_{1k}(\tau)  \approx  - \frac{3}{8 \pi^2} \frac{ \alpha_o \Omega  \gamma^2 }{\Omega_P^4}  \int_{0}^{\tau} d\lambda \ \int_0^\infty d k \int_0^1 dy \   \sin \Omega (\tau - \lambda) k^6   \coth(\beta k/2)  y^2   \ e^{ - \frac{\gamma k^2}{k^2 + \Omega_P^2} (\tau + \lambda) -  2 k y z}
\end{eqnarray}
where a more precise comparison of the thermal wavelength to the time and distance scales is necessary before we can expand the hyperbolic cotangent.
Note that the combination of exponentials in the integrand ensures convergence for the $k$-integration and that replacing $- \frac{\gamma k^2}{k^2 + \Omega_P^2} (\tau + \lambda) \rightarrow - \frac{\gamma k^2}{ \Omega_P^2} (\tau + \lambda)$ results in a negligible error when $ \gamma \tau >> 1$. Making this replacement allows us to perform the $y$-integration which to leading order for large $z$ gives
\begin{eqnarray}
\label{ }
f^{dyn}_{1k}(\tau)  \approx  - \frac{3}{32 \pi^2} \frac{ \alpha_o \Omega  \gamma^2 }{\Omega_P^4} \frac{1}{z^3}   \int_{0}^{\tau} d\lambda \ \int_0^\infty d k \   \sin \Omega (\tau - \lambda) k^3 \coth(\beta k/2)  e^{ - \frac{\gamma k^2}{ \Omega_P^2} (\tau + \lambda) }. 
\end{eqnarray}

At room temperature (or hotter) time is the factor which determines the largest scale in the problem when  $\tau \gtrsim 1 \text{ns}$ (for a gold surface), thereby allowing a Taylor expansion of the hyperbolic cotangent in the integrand. Subsequently the $k$ and $\lambda$ integral can be performed where the result below is expressed to leading order for $\Omega \tau >>1$

\begin{eqnarray}
\label{fdyn_app}
f^{dyn}_{1k}(\tau)  \approx  - \frac{3  \alpha_o }{16 \pi^{3/2} } \sqrt{    \frac{\gamma}{ \Omega_P^2 } }  \frac{ T}{z^3}  \bigg[ \frac{1}{2 \sqrt{2}}  - \cos \Omega \tau
\bigg] \frac{1}{  \tau^{3/2}}. 
\end{eqnarray}
 A few remarks are in order. First, it is not surprising that the force depends on the inverse of the cubic distance. In a nonequilibrium steady-state where the field and matter are described by different temperatures it is an evanescent field that gives rise to the novel scaling of $1/z^3$ in the far-field \cite{Ant_JPhys}. The particular case we address here is the time-dependence of the evanescent field as it is thermalized by the eddy currents. 
Second, as the conductivity becomes very good $\Omega_P/\gamma >>1$ the presence of eddy currents is suppressed (in this limit the Drude model approaches plasma model). 
This effect is captured by the prefactor $\sqrt{ \gamma/ \Omega_P^2 }$ in (\ref{fdyn_app}) which relates to the resistivity of the material indicating that the force is larger for poor conductors. 

The plots in Figure[2] quantify (\ref{fdyn_app}) for a gold surface and a rubidium atom when the field is initially at room temperature. 
\begin{figure}
\centering
\mbox{\subfigure{\includegraphics[width=3in]{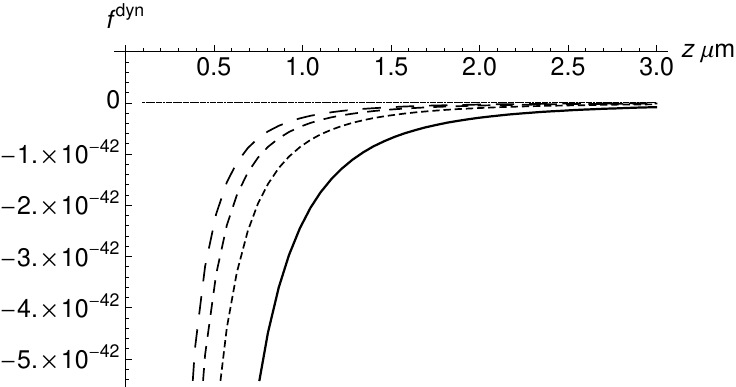}}\quad
\subfigure{\includegraphics[width=3in]{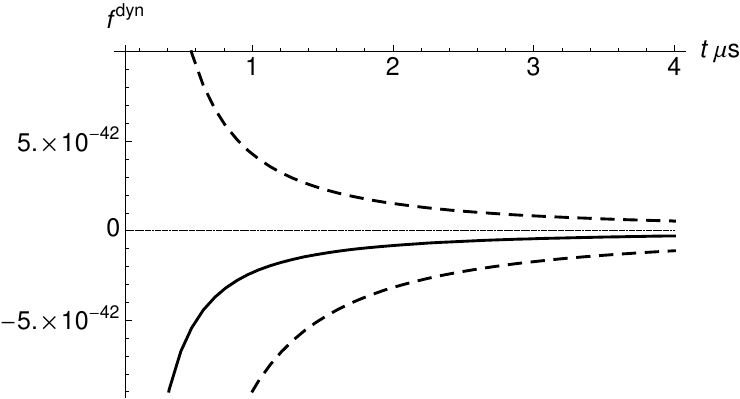} }}
\caption{The plots above quantify the dynamical force (in Newtons) at room temperature (T = 295 K) between a rubidium atom and a gold surface as the field is thermalized by eddy currents. The plasma frequency and dissipation rate for gold used were $\Omega_p = 8.9$ eV and $\gamma = 0.0357$ eV. For rubidium we set the static polarizability, $\alpha_o$, equal to $ 4.73 \times 10^{-29} m^3$ and $\Omega$ to rubidium's first optical resonance $2.35 \times 10^{15}$ Hz. \textit{The plot on the left} shows the spatial dependence of the atom-surface force (neglecting the rapidly oscillating  cosine term in (\ref{fdyn_app})) as a function of distance in microns evaluated at different times. The solid line corresponds to one microsecond. From there each successive line moving upward gives the force at one microsecond intervals from $1 \mu s$ to $ 4\mu s$ for the top line. \textit{The plot on the right} shows the time dependence of the dynamical force as a function of time at an atom surface distance of one micron. The force is rapidly oscillating and so we've plotted the average force, indicated by the solid line, and the envelope, by the dashed lines, bracketing the force oscillations. 
} \label{fig12}
\end{figure}

%%%%%%%%%%%%%%%%%%%%%%%%%%%%%%%
\section{Conclusion}
%%%%%%%%%%%%%%%%%%%%%%%%%%%%%%%

In this paper we have derived from first principles with a microscopic model the nonequilibirum force between an atom and a
medium (modeled by a continuous lattice of non-interacting harmonic oscillators each coupled to a reservoir) including fully dynamical
activities such as dissipation, absorption, radiation and fluctuations.
In the end we show that the force is in
agreement with the results of previous works employing MQED
theory when the long-time limit is taken and the system has evolved to a steady state.

Our results go beyond previous work by providing the fully dynamical 
force including effects of relaxation, thermalization, and backaction from the field and matter
in addition to being valid for general dielectric geometries and composition.
These new results could play a role in experiments measuring the dynamical aspects of 
surface atom interactions, like quantum friction \cite{Volokitin08, Philbin09} (and references therein) and relaxation. The dynamical effects 
derived herein are very sensitive to the model used to describe the dielectric. In particular
one would expect stark differences between the dynamics of the force when comparing 
Ohmic models, like Drude, with the plasma model which neglects dissipation. Dynamical effects 
not only provide a new arena in which to explore quantum fluctuation forces but could aid in the 
resolution of controversies involving quantum friction \cite{Volokitin09,Pendry10,Leonhardt10} 
and the so-called thermal problem \cite{Milton04,Klimchitskaya09}.

As a final note we point out that beyond giving a fully dynamical description
for atom motion in the presence of a dielectric material the nonequilibrium
quantum field theoretic formulation of this problem is particularly adept at
treating fluctuation phenomena. It will be the aim of a future study to understand the
rich interplay between the propagating and evanescent wave components
of the force and their effect upon the fluctuations of the atom's trajectory in space.

\section{Appendix}

%%%%%%%%%%%%%%%%%%%%%%%%%%%%%%%
\subsection{Evanescent and Propagating Waves}
%%%%%%%%%%%%%%%%%%%%%%%%%%%%%%%

To understand the nature of the nonequilibrium force correction (\ref{fneq})
 we study the Green's function $\mathcal{G}^{ret}_{ik}({\omega'}, \vec{x}, \vec{y} )$
 in the vacuum region.
 In equilibrium the fluctuations of the field in this region ($z > 0$) come in two categories; left-incident
 waves and their reflected components, and waves that partially transmit into the vacuum region. By 
 use of the fluctuation-dissipation relation we will use  $\mathcal{G}^{ret}_{ik}({\omega'}, \vec{x}, \vec{y} )$
 to identify the effects associated with partially transmitted field fluctuations. 
 
In the vacuum half-space
 $\mathcal{G}^{ret}_{ik}({\omega'}, \vec{x}, \vec{y} )$
  is obtained by solving the classical equations of motion for the
field, subject to dielectric boundary conditions in the $z=0$ plane,
sourced by a delta function. Its form can be taken from \cite{Tomas95} or the Appendix
of \cite{APSS08} and is written below for both spatial arguments
lying in the vacuum half-space:

\begin{equation}
\label{ }
\mathcal{G}^{ret}_{ij}({\omega'}, \vec{x},\vec{x}')=\mathcal{G}^o_{ij}+ \frac{ i
{\omega'}^2}{  8 \pi^2 k_z}  \sum_{\sigma=TE,  TM} \int d^2 k_\| \
e_{\sigma,i}(+) e_{\sigma,j}(-) R_{\sigma} e^{i k_z(z+z')} e^{i
\vec{k}_\| \cdot (\vec{x}_\|- \vec{x}'_\|)}.
\end{equation}
Above $R_{\sigma}$ are the Fresnel reflection coefficients where the
index $\sigma$ refers to the polarization (implicitly appearing below (\ref{RTE})
and (\ref{RTM})). The polarization vectors
for the field in the vacuum region are $e_{\sigma,i}(\pm)$ where $e_{TE}(\pm)=\hat{k}_\| \times
\hat{z}$ and $e_{TM}(\pm)=(k_\| \hat{z}\mp  k_z \hat{k}_\|)/{\omega'}$,
$\hat{k}_\|$ and $\hat{z}$ are unit vectors, $\hat{z}$ normal to
the vacuum dielectric interface with  $\hat{k}_\|$ directed
parallel to the projection of the wave vector in the plane of the
interface, and $k_z = \sqrt{{\omega'}^2-k_\|^2}$ is the z-component of
the wavevector.  The free field component of the Green's functions, or bulk term, is $\mathcal{G}^o_{ij}$
and would be present whether the dielectric were there or
not. The bulk term leads to a divergence in the energy but has 
no spatial dependence at coincidence, and so we discard
it as we require the force to vanish at infinite atom-surface
spacing.

Let us now focus on the portion of the atom-surface force in thermal equilibrium that arises from  
fluctuations of the field, be them induced or intrinsic given by % from now on the PF-contribution as

\begin{eqnarray}
\label{ }
f^{FF}_k  &
=  &  \frac{q^2}{4\pi} \int_{-\infty}^{\infty} d{\omega'}
\delta^{ij}   g^*_{ret}({\omega'}) \partial_k(x)
\mathcal{G}^{H}_{ij}({\omega'}, \vec{x},\vec{z})
\bigg|_{\vec{x}=\vec{z}(\tau)}
\nonumber \\
&
=  &   2 \int_{-\infty}^\infty d{\omega'} \ \alpha^*({\omega'}) \coth[\beta {\omega'}/2 ]  \delta^{ij} \partial_k(x)
 \text{Im} [\mathcal{G}^{ret}_{ij}({\omega'}, \vec{x},\vec{z}) ]
\bigg|_{\vec{x}=\vec{z}(\tau)}
\end{eqnarray}
where $FF$ stands for field fluctuations.
Tracing over the indices of the Green's function, taking the spatial derivative
 and setting the two spatial arguments to the
position of the atom we can express the FF-contribution to the force as

\begin{equation}
\label{fpf}
f^{FF}_{k} =-\frac{1}{4\pi^2}  \int_{-\infty}^\infty d{\omega'}  \int d^2 k_\| \  {\omega'}^2 \alpha^*({\omega'})  \coth(\beta {\omega'}/2) \text{Im} \bigg\{ \bigg[ R_{TE}+ R_{TM} \bigg( \frac{k_\|^2-k_z^2}{{\omega'}^2} \bigg) \bigg] e^{i2 k_z z} \bigg\}.
\end{equation}
 When both
 spatial arguments lie in the vacuum half-space $\mathcal{G}^{ret}_{ik}({\omega'}, \vec{x}, \vec{y} )$
 decomposes into two contributions.
 First, field fluctuations in the
vacuum region give rise to propagating waves, in particular, waves
moving in the negative $z$-direction will reflect from the dielectric
surface giving rise to waves propagating in the positive
$z$-direction. Second, a field fluctuation within the dielectric
producing waves  propagating toward the vacuum-dielectric interface
will partially transmit into the vacuum region. One consequence of this
is that the $z$-component of the wave vector in (\ref{fpf}) is not
necessarily real.  For values of $|k_\|| < {\omega'}$ we see that $k_z$
is real but becomes pure imaginary for the integration range $|k_\||
\in ({\omega'}, \infty)$. The former is associated with the propagating
solutions to the wave equation in the vacuum and the latter with
evanescent waves.

We can isolate the influence of the evanescent modes on the force 
if we restrict the integration range so that the magnitude of the 
transverse momenta are strictly greater than the wave frequency

\begin{equation}
\label{}
f^{FF, EW}_{k}=-\frac{1}{\pi}  \int_0^\infty d{\omega'}  \int_{\omega'}^\infty d k_\| k_\| \ {\omega'}^2 \text{Re}[ \alpha({\omega'}) ] \coth(\beta {\omega'}/2) \text{Im} \bigg\{ \bigg[ R_{TE}+ R_{TM} \bigg( \frac{k_\|^2-k_z^2}{{\omega'}^2} \bigg) \bigg] e^{i2 k_z z} \bigg\}
\end{equation}
where $EW$ stands for evanescent waves.

In this range the $z$-component of the wave vector is pure imaginary, 
$k_z\rightarrow i\kappa$, and this property can be used to simplify the 
equation for the force by explicitly taking the imaginary part of the 
integrand inside the curly brackets $\{..\}$.
We note that

\begin{equation}
\label{RTE}
\text{Im} [R_{TE} ]=\text{Im} \bigg[\frac{k_z-k_z'}{k_z+k_z'} \bigg]= \frac{2\kappa \text{Re}[k_z'] }{|k_z+k_z'|^2}
\end{equation}

\begin{equation}
\label{RTM}
\text{Im} [R_{TM} ]=\text{Im} \bigg[\frac{\varepsilon({\omega'}) k_z-k_z'}{\varepsilon({\omega'}) k_z+k_z'} \bigg]= \frac{2\kappa \text{Re}[\varepsilon({\omega'}) k_z^{'*}] }{|\varepsilon({\omega'}) k_z+k_z'|^2}= \frac{2\kappa \text{Re}[ k_z^{'}]  (k_\|^2+|k_z'|^2)}{{\omega'}^2 |\varepsilon({\omega'}) k_z+k_z'|^2}
\end{equation}
where $k_z'=\sqrt{\varepsilon({\omega'}) {\omega'}^2-k_\|^2}$ is the $z$-component of the wave vector in the dielectric and in the last step we have used the identity ${\omega'}^2 \text{Re}[\varepsilon({\omega'}) k_z^{'*}] =  \text{Re}[ k_z^{'}]  (k_\|^2+|k_z'|^2)$.
To bring the force due to evanescent waves into its final form we make the change of variables $k_\|={\omega'} q$ and note that $\sqrt{2}\text{Re}[k_z']= \sqrt{\text{Re}[\varepsilon({\omega'})]{\omega'}^2-k_\|^2+|\varepsilon({\omega'}){\omega'}^2-k_\|^2|}$

\begin{eqnarray}
\label{PFEWc}
f^{FF, EW}_{k} =-\frac{\sqrt{2}}{\pi}  \int_0^\infty d{\omega'}  \int_1^\infty d q q \  {\omega'}^4 \text{Re}[ \alpha({\omega'}) ] \coth(\beta {\omega'}/2) \sqrt{q^2-1}
 \sqrt{\text{Re}[\varepsilon({\omega'})]-q^2 +|\varepsilon({\omega'})-q^2|} \nonumber \\
\times \bigg[
\frac{1}{|\sqrt{1-q^2}+\sqrt{\varepsilon({\omega'})-q^2} |^2} +\frac{(2q^2-1)(q^2 -|\varepsilon({\omega'})-q^2|)}{|\varepsilon({\omega'}) \sqrt{1-q^2}+\sqrt{\varepsilon({\omega'})-q^2} |^2}
 \bigg] e^{-2 z {\omega'} \sqrt{q^2-1} }.
\end{eqnarray}
This expression now quantifies the force on the atom due only to evanescent field fluctuations. 

%%%%%%%%%%%%%%%%%%%%%%%%%%%%%%%
\subsection{Nonequilibrium Correction: Relation to the Equilibrium Force}
%%%%%%%%%%%%%%%%%%%%%%%%%%%%%%%

In this section we explicitly evaluate the  the nonequilibrium correction 
to the force beginning with the convolution of the Green's functions in $f^{neq}_k$.
Starting from (\ref{fneq}) we note that the Green's functions
 appearing are those with support at one
point in the vacuum half-space and the other located within the
medium. Using the appendices of the papers 
\cite{Henkel} and \cite{Mara} we can evaluate the convolution of the Green's function
product

\begin{eqnarray}
\label{ }
I & \stackrel{def}{=}  & \int_{-\infty}^{\infty} dx \int_{-\infty}^{\infty} dy \int_{-\infty}^{0} dz \  \mathcal{G}^{ret,L}_{ij}({\omega'}, \vec{r}_1,\vec{x}) \partial_z \mathcal{G}_{ret,L}^{ij*}({\omega'}, \vec{r}_2, \vec{x})  \\ 
& = & \frac{1}{16\pi^2} \int d^2k_{\|} e^{i \vec{k}_{\|} \cdot (\vec{r}_{1\|}-\vec{r}_{2\|})} \frac{(- i k_z^*)  {\omega'}^4}{2 |k_z'|^2 \text{Im}[k_z']}  \bigg[ |T_{TE}|^2 +|T_{TM}|^2 \left( \frac{ k^2_\|+|k_z|^2}{{\omega'}^2} \right) \left( \frac{ k^2_\| +  | k_z'|^2}{|\varepsilon({\omega'}) | {\omega'}^2} \right) \bigg]
e^{i (k_z z_1- k_z^* z_2)}.
 \nonumber
\end{eqnarray}
Above $T_{TE}$ and  $T_{TM}$ are the Fresnel transmission coefficients for the TE and TM-polarized waves. After making the change of variables $k_\|= {\omega'} q $, plugging in the explicit forms for the Fresnel coefficients, and setting $\vec{r}_1$ and  $\vec{r}_2$ to be the position of the atom $I$ separates into two contributions, one from evanescent waves and the other propagating waves (labeled $EW$ and $PW$)

\begin{eqnarray}
\label{ }
I_{EW}
 =
  -\frac{\sqrt{2} }{4 \pi} \int_{1}^\infty dq q  \frac{{\omega'}^3 |{\omega'}| \sqrt{q^2-1}}{ \text{Im}[\varepsilon({\omega'})]}   \sqrt{\text{Re}[\varepsilon({\omega'})]-q^2 +|\varepsilon({\omega'})-q^2|}  e^{-2z | {\omega'} | \sqrt{q^2-1}}
 \nonumber \\
 \times
 \bigg[ \frac{1}{|\sqrt{1-q^2}+\sqrt{\varepsilon({\omega'})-q^2}
 |^2} +\frac{(2q^2-1)(q^2 + |\varepsilon({\omega'})-q^2|)}{|\varepsilon({\omega'})
\sqrt{1-q^2}+\sqrt{\varepsilon({\omega'})-q^2} |^2}  \bigg]
  \nonumber
\end{eqnarray}

\begin{eqnarray}
\label{ }
I_{PW}
 =
  -\frac{ i \sqrt{2} }{4 \pi} \int_{0}^1 dq q  \frac{{\omega'}^4 \sqrt{1-q^2}}{ \text{Im}[\varepsilon({\omega'})]}   \sqrt{\text{Re}[\varepsilon({\omega'})]-q^2 +|\varepsilon({\omega'})-q^2|}  
 \nonumber \\
 \times
 \bigg[ \frac{1}{|\sqrt{1-q^2}+\sqrt{\varepsilon({\omega'})-q^2}
 |^2} +\frac{(q^2 + |\varepsilon({\omega'})-q^2|)}{|\varepsilon({\omega'})
\sqrt{1-q^2}+\sqrt{\varepsilon({\omega'})-q^2} |^2}  \bigg].
  \nonumber
\end{eqnarray}
For evanescent waves we have constrained the imaginary wave vector $k_z$ to be positive by expressing it as $k_z = i |{\omega'}| \sqrt{q^2-1}$ after the change of variables $k_\| \rightarrow {\omega'} q$. 
Combining this with the expression for the nonequilibrium correction to the force (\ref{fneq}) we can express the force as (\ref{fneq_ew}) and (\ref{fneq_pw}).

%It is interesting to note that the atom-surface force computed with respect to the initially factorized state contains an additional evanescent component arising from fluctuations of the medium.

%%%%%%%%%%%%%%%%%%%%%%%%%%%%%%%
\acknowledgments

R. B. is indebted to Francesco Intravaia for many useful discussions and would like to thank Felipe da Rosa for providing several thoughtful suggestions.
This work is supported in part by the Maryland Center for Fundamental Physics and by NSF grants PHY-0426696 and PHY-0801368 and partially funded by DARPA/MTOÕs Casimir Effect Enhancement program under DOE/NNSA Contract
DE-AC52-06NA25396

%%%%%%%%%%%%%%%%%%%%%%%%%%%%%%%


\begin{references}
\bibitem{FN1}
We should point out here that when the medium and the field interact very strongly the assumption that the initial state is factorizable may not be justified.  We adopt such a state here for ease of computation.  For more general situations one can introduce a preparation function to describe the initial state,  but as shown in \cite{Romero97}
the key features are not affected too strongly.
\bibitem{FN2}
arbitrary composition means that we are free to choose the spectral density of the reservoir
\end{references}
\end{document}